\documentclass[10pt,twocolumn,letterpaper]{article}

\usepackage{cvpr}              
\usepackage{graphicx}
\usepackage{amsmath}
\usepackage{amssymb}
\usepackage{booktabs}
\usepackage{tikz}
\usepackage{pgfplots}
\usepackage{tablefootnote}
\usepackage{xcolor}
\usepackage{pgfplotstable}
\usepgfplotslibrary{colorbrewer}
\pgfplotsset{compat=1.17}
\usepackage{adjustbox} 
\usepackage{booktabs} 
\definecolor{purple}{rgb}{0.65,0,0.65}
\definecolor{dark_green}{rgb}{0, 0.5, 0}
\definecolor{blueish}{rgb}{0.0, 0.3, .6}
\definecolor{LightCyan}{rgb}{0.88,0.95,1}
\definecolor{tabhighlight}{HTML}{e5e5e5}
\usepackage{siunitx}
\usepackage{subcaption}
\usepackage{array}
\usepackage{multirow} 
\usepackage{soul}
\newcommand{\EmbedAttack}{Ours $\left(\mathcal{L}_{emb} \right)$\@\xspace}

\newcommand{\DepthPGD}{\textsl{DepthPGD}\@\xspace}
\newcommand{\SegPGD}{\textsl{SegPGD}\@\xspace}

\newcommand{\AttnAttackVisual}{Ours $\left (\mathcal{L}_{atn}\right)$\@\xspace}

\newcommand{\CombAttack}{Ours $\left(\mathcal{L}_{comb}\right)$\@\xspace}

\definecolor{cvprblue}{rgb}{0.21,0.49,0.74}
\usepackage[pagebackref,breaklinks,colorlinks,allcolors=cvprblue]{hyperref}

\def\paperID{7} 
\def\confName{CVPR}
\def\confYear{2025}

\title{Attacking Attention of Foundation Models Disrupts Downstream Tasks}

\author{Hondamunige Prasanna Silva\\
University of Florence\\
{\tt\small hondamunige.silva1@edu.unifi.it}
\and
Federico Becattini\\
University of Siena\\
{\tt\small federico.becattini@unisi.it}
\and
Lorenzo Seidenari\\
University of Florence\\
{\tt\small lorenzo.seidenari@unifi.it}
}

\begin{document}
\maketitle

\begin{abstract}
Foundation models represent the most prominent and recent paradigm shift in artificial intelligence.
Foundation models are large models, trained on broad data that deliver high accuracy in many downstream tasks, often without fine-tuning. For this reason, models such as CLIP \cite{radford2021learning}, DINO \cite{oquab2023dinov2} or Vision Transfomers (ViT) \cite{dosovitskiy2020image}, are becoming the bedrock of many industrial AI-powered applications. 
However, the reliance on pre-trained foundation models also introduces significant security concerns, as these models are vulnerable to adversarial attacks. Such attacks involve deliberately crafted inputs designed to deceive AI systems, jeopardizing their reliability.
This paper studies the vulnerabilities of vision foundation models, focusing specifically on CLIP and ViTs, and explores the transferability of adversarial attacks to downstream tasks. We introduce a novel attack, targeting the structure of transformer-based architectures in a task-agnostic fashion.
We demonstrate the effectiveness of our attack on several downstream tasks: classification, captioning, image/text retrieval, segmentation and depth estimation.

\end{abstract}    
\section{Introduction}\label{sec:intro}

Foundation models are nowadays powering the vast majority of AI-based algorithms and technologies. Foundation models are large models that are trained on broad data and are easily fine-tuned to downstream tasks. A trend started with the first ``off-the-shelves" image recognition models pre-trained on Imagenet~\cite{sharif2014cnn}, is now brought forward by remarkable advancements by large language models~\cite{devlin2018bert, brown2020language}, vision backbones~\cite{oquab2023dinov2} and multimodal models~\cite{radford2021learning}. 

Among the foundation models, CLIP~\cite{radford2021learning} can be regarded as one of the most prominent players in all modern computer vision applications.
It relies on contrastive learning to learn an optimal alignment between textual and visual embeddings, and it is widely used in downstream tasks including classification, retrieval and captioning~\cite{li2023blip}. 
Like most foundation models, CLIP relies on attention, which weighs different parts of the input and also plays a role in spatial information handling.
Similarly, pretrained DINO models \cite{oquab2023dinov2} are widely used as feature extractors, especially for fine-grained tasks such as segmentation or depth estimation.
\begin{figure}
    \includegraphics[width=\linewidth]{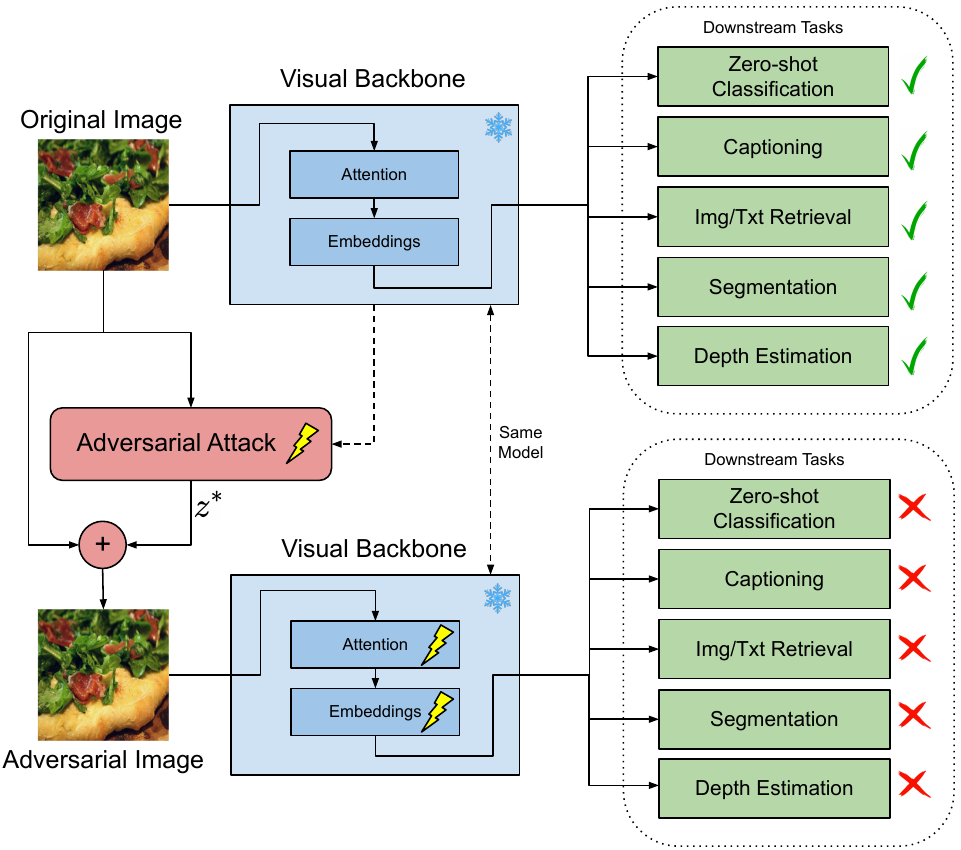}
    \caption{Our method generates adversarial noise $z^*$ by attacking attention and embeddings in visual backbones of foundation models, without any knowledge of downstream tasks. Unlike other attacks, access to other modalities (e.g., text) is not required.} 
    \label{fig:eye-catcher}
\end{figure}
However, foundational models, like all deep learning models, are vulnerable to adversarial attacks. In fact, the widespread application of single models throughout an array of different tasks makes attacks potentially more dangerous, as a single attack could affect multiple applications.
Nonetheless, attacks on foundation models are currently successfully performed by exploiting knowledge about individual downstream tasks \cite{lu2023set}.
When the foundational model is multimodal, e.g. image and text, joint attacks on both modalities can also be conducted~\cite{wang2023exploring} to improve their effectiveness. However, this requires access to both modalities, making the attacks not always feasible in practice.

In this work, we propose an attack for foundation models, specifically focusing on CLIP and Vision Transformer backbones, by taking a different angle on the problem.
Unlike existing methodologies, which are often multimodal and task-specific, we design our attack to target only the visual domain
and to be agnostic to the downstream task.
In fact, we believe that to make the attack more disruptive, foundational models should be attacked independently from the downstream tasks.
Our method, by jointly damaging attention and the final model embeddings, can attack with the same perturbation all downstream applications, as shown in Fig.~\ref{fig:eye-catcher}.
We show that tasks like zero-shot classification, retrieval, captioning, segmentation and depth estimation are all affected by our simple attack. These results highlight the potential fragility of foundation models. Moreover, attacks in the image domain are harder to spot especially when conducted with low budgets, compared to attacks that perturb both images and text.
Interestingly, attacks targeting specific parts of attention layers have different impacts, depending on the nature of the downstream task: fine-grained tasks, as captioning or segmentation/depth estimation, are better addressed by disrupting visual attention, whereas more semantic tasks, as classification and retrieval, are better targeted by attacks on embeddings.
In summary, our main contributions are the following:
\begin{itemize}
    \item We propose an effective adversarial perturbation algorithm targeting attention of vision backbones in foundational models.
    \item Our method is task agnostic, i.e. can attack samples without their label or textual description.
    \item Extensive experimentation shows that our method can affect downstream tasks such as captioning, classification, retrieval, segmentation and depth estimation, reporting state-of-the-art results in terms of attack success rate.
\end{itemize}

\section{Related works}
\paragraph{Vision Foundation Models}

Vision Foundation Models, have become pivotal in advancing multimodal AI by effectively addressing downstream tasks across various domains. These models, trained on large-scale image-text datasets, learn representations that can be fine-tuned for specific tasks, significantly improving performance.
Whereas there are examples of purely visual foundation models \cite{dosovitskiy2020image,oquab2023dinov2,kirillov2023segment}, most of these models are multi-modal, exploiting a Vision-Language Pre-training (VLP) that jointly learns to process images and text.
CLIP \cite{radford2021learning} is a landmark model in this field, enabling zero-shot open-vocabulary image classification by aligning images with textual descriptions. This model set a new benchmark for generalization in vision-language tasks. Following CLIP, ALIGN \cite{jia2021scaling} scaled this approach, improving the performance on tasks like image retrieval and cross-modal retrieval by leveraging billions of image-text pairs.
Florence \cite{yuan2021florence} exemplifies the application of VLPs in multi-task learning. Trained on a vast dataset, it excels in object detection, semantic segmentation, and image captioning, demonstrating the benefits of joint vision-language optimization.
Regarding purely visual backbones, DINO \cite{caron2021emerging} and its improved version DINOv2 \cite{oquab2023dinov2}, introduced a self-supervised approach to train Vision Transformers, yielding extremely versatile backbones that can be used across tasks even without fine-tuning.
In this work, we focus on attacking CLIP and Vision Transformer backbones as they are largely employed as core building blocks of a plethora of downstream applications.

\paragraph{Adversarial Attacks on ViT}
With the extensive use of Vision Transformers (ViTs)~\cite{dosovitskiy2020image} as the core of various vision foundation models \cite{radford2021learning,kirillov2023segment}, there is a growing interest in evaluating their robustness to pinpoint their weaknesses. The Pay No Attention (PNA)~\cite{wei2022towards} method adapts the Skip Gradient Method (SGM) for ViTs, specifically omitting the gradient of the attention block during back-propagation to enhance the transferability of adversarial examples through gradient regularization. The attack also introduces the PatchOut strategy, which randomly selects a subset of patches to compute the gradient in each attack iteration, serving as an image transformation technique for transferable adversarial attacks on ViTs.
Another notable method is the Token Gradient Regularization (TGR)~\cite{zhang2023transferable}:  leveraging the structural characteristics of ViTs, TGR reduces the variance of back-propagated gradients on a token-wise basis and utilizes the regularized gradient to generate adversarial samples. Contrasting with these strategies, Naseer et al.~\cite{naseer2021improving} introduced the Self-Ensemble (SE) method and the Token Refinement (TR) module to boost the transferability of adversarial examples generated by ViTs. SE leverages the classification token at each ViT layer with a shared classification head for feature-level attacks. Building on SE, TR further refines the classification token through fine-tuning to improve attack performance. 

In general, the literature presents conflicting conclusions regarding the adversarial robustness of transformers, compared to the one of Convolutional Neural Networks (CNNs). One school of thought posits that transformers demonstrate superior robustness compared to CNNs. Recent studies~\cite{kim2022analyzing, shao2021adversarial, aldahdooh2021reveal} attribute this robustness to the differing reliance on input features. Specifically, CNNs depend heavily on high-frequency information, whereas transformers do not. Consequently, transformers are believed to be more resilient to gradient-based attacks like Fast Gradient Sign Method (FGSM)~\cite{goodfellow2014explaining}, Projected Gradient Descent (PGD)~\cite{pgd}, and Carlini \& Wagner (C\&W) \cite{carlini2017towards} attacks.
Furthermore, the severe nonlinearity inherent in the input-output relationships of transformers may contribute to their enhanced robustness \cite{kim2024curved}. In scenarios involving adversarial training, Vision Transformers (ViTs) exhibit superior generalization and robustness compared to CNNs, as highlighted by Liu et al. \cite{liu2022exploring}.

Contrarily, another body of research contends that transformers are just as susceptible to adversarial attacks as CNNs. The findings of Mahmood et al. \cite{mahmood2021robustness} indicate that ViTs do not outperform ResNet architectures in terms of robustness against various attack methods including FGSM~\cite{goodfellow2014explaining}, PGD~\cite{pgd}, and C\&W~\cite{carlini2017towards}. Similarly, there is evidence that both CNNs and transformers exhibit comparable vulnerabilities to natural and adversarial perturbations~\cite{bhojanapalli2021understanding}, and a direct comparison under a unified training setup revealed that both architectures possess similar levels of adversarial robustness \cite{bai2021transformers}. Additional vulnerabilities of Vision Transformers are exposed through targeted patch attacks \cite{fu2022patch,gu2022vision}.
Recently, EmbedAttack \cite{kowalczuk2024benchmarking} was proposed as the first task agnostic attack on ViT backbones, by simply applying PGD on the embeddings rather than on the targets.
In this work, differently from EmbedAttack, we target ViT backbones of foundation models by attacking its core component: attention. Our attack crafts adversarial noise to make the model focus on irrelevant parts of the image, yielding performance drops across several downstream tasks, without knowing a-priori which task the backbone could be used for.
In addition, we also show that attacking different parts of the attention layer (e.g., the attention matrix of the final embedding), yields different degrees of efficacy depending on the nature of the downstream task.

\paragraph{Attacks on Multimodal Foundation Models}

Unlike standard attacks, which typically target the task by exploiting the relationship between a single instance and its label, attacks on vision-language models must account for multi-modal data relationships. Fort \cite{Fort2021CLIPadversarial} demonstrated that ``standard" adversarial attacks are effective on CLIP models when customized for a particular downstream task. Other studies \cite{Fort2021CLIPadversarialstickers,noever2021reading} examined attacks on multi-modal CLIP neurons by placing text stickers on images, prompting the CLIP model to ``read" the text and ignore the rest of the image. These attacks adhere to a different threat model compared to traditional budget-based attacks, as the adversarial manipulations are quite evident. Notably, \cite{carlini2023poisoning} recently revealed that poisoning attacks on web-scale datasets are feasible and can significantly affect the training process of foundational models like CLIP.
Zhang et al. \cite{zhang2022towards} introduced the Collaborative Multimodal Adversarial Attack (Co-Attack) to target various pre-trained models, such as CLIP, ALBEF~\cite{li2021align}, and TCL~\cite{yang2022vision}. Co-Attack generates multi-modal adversarial perturbations by increasing the embedding distance between adversarial examples and their original data pairs.
The Set-level Guidance Attack (SGA) \cite{lu2023set} uses data augmentation to boost data diversity, thereby creating multi-modal adversarial perturbations by minimizing the similarity between them and their matched data from another modality. Additionally, the Self-augment-based Transfer Attack (SA-Attack) \cite{he2023sa} applies further data augmentations to both the original data and the adversarial examples to improve attack transferability across models. While Co-Attack, SGA, and SA-Attack typically generate adversarial perturbations sequentially, first for one modality and then for another, methods for learning multi-modal adversarial perturbations simultaneously also exist~\cite{wang2023exploring}.
All aforementioned attacks depend on task-specific characteristics, hindering their broader applicability.
In contrast, our proposed method makes a significant advancement by adopting a task-agnostic approach. It simultaneously disrupts both the attention mechanism and the image embedding within attention-based models, without relying on specific task labels or characteristics.
This dual disruption effectively shifts the embedding away from its correct representation and manipulates the attention distribution across the input, making it applicable to a wide range of downstream tasks.
By targeting the foundational backbone rather than the task-specific components, our method provides a more robust, scalable, and universally applicable adversarial strategy.
It also simplifies the attack pipeline on multimodal models, by solely attacking the vision branch. Whereas leveraging a joint attack could yield more effective disruptions, not accessing other modalities greatly improves the attack applicability, as no additional information on the data other than the image is required.

\begin{figure}[t]
    \includegraphics[width=0.45\textwidth]{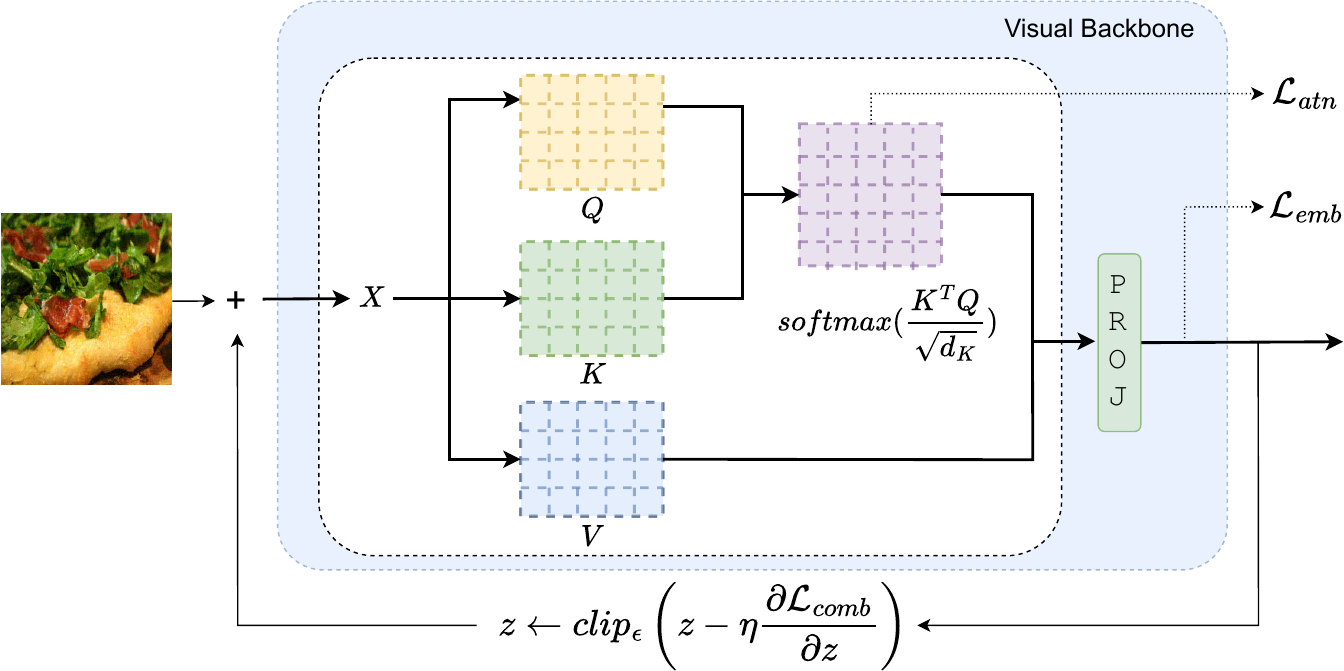}
	\caption{Our attack disrupts both the attention mechanism and image embedding in attention-based models.}
	\label{fig:methodology}
\end{figure}

\section{Attacking Foundation Models}
We introduce a novel adversarial attack that simultaneously targets the attention mechanism and image embedding in attention-based models, as shown in Fig. \ref{fig:methodology}. The core objective of our approach is to disrupt the input's representation by shifting it away from its original embedding and altering how the model distributes attention across different parts of the input. 
Although simple, what makes our attack strategy appealing is that it is completely agnostic to ground truth labels and, since it is tailored to attack foundational backbones, it is also agnostic to the nature of any downstream task. 
Our proposed attack is versatile and can be applied to any vision transformer or attention-based model. However, motivated by its agnostic nature, we focus on foundational models, which are frequently leveraged either by tuning only task-specific heads or in zero- or few-shot settings.
The main idea of our method is an attack that can target said backbones, and by damaging the original model, also disrupts the performance of any downstream task that builds upon its learned representations.

\subsection{Optimizing the Perturbation}

Given a model \( f(x) \) with input \( x \), our objective is to generate a perturbation \( z \) such that the adversarial example \( x' = x + z \) causes the model output to diverge from its intended outcome. To achieve this, we optimize the perturbation \( z \) by considering the gradients of both the attention mechanism and the embedding space with respect to the input. The optimization problem can be formally defined as

\begin{equation}
z^* = \arg\min_{z} \mathcal{L}_{comb}(f(x + z))
\end{equation}

where \( \mathcal{L}_{comb} \) is a combined loss that integrates both attention and embedding losses, as described further. The perturbation is iteratively updated using gradient descent:

\begin{equation}
z \leftarrow \text{clip}_{\epsilon} \left( z - \eta \frac{\partial \mathcal{L}_{comb}}{\partial z} \right)
\end{equation}

where \( \eta \) is the learning rate, computed using Weighted Adam \cite{loshchilov2017decoupled} and $\epsilon$ is the perturbation budget.
The function $\text{clip}_{\epsilon}(\cdot)$ is the clipping function that ensures that adversarial noise does not exceed the budget at every iteration, that is, $\left\| z \right\|_{\infty } < \epsilon$.
The optimization alters the embedding of the perturbed input while simultaneously altering the attention distribution. 

\begin{figure}[!t]
    \centering
    \begin{minipage}[t]{0.4\columnwidth}
        \centering
        \includegraphics[width=1\linewidth]{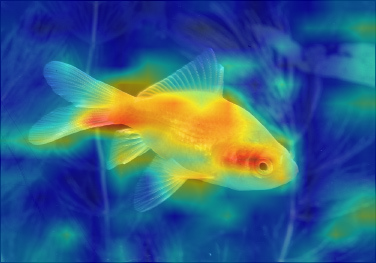} 
    \end{minipage}~~~
    \begin{minipage}[t]{0.4\columnwidth}
        \centering
        \includegraphics[width=1\linewidth]{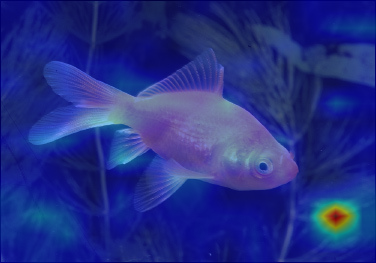} 
    \end{minipage}
    \caption{Attention on clean (left) and perturbed (right) image.\label{fig:side_by_side}}
    
\end{figure}

\subsection{Attention Manipulation}
Our approach manipulates the attention matrix in transformers to redirect the model's focus from significant tokens to less relevant ones, thereby impairing its performance, as shown in Fig.~\ref{fig:side_by_side}. The attention mechanism in transformers projects the input into three embedding vectors in $\mathbb{R}^{d_k}$, the query \textbf{Q}, the key \textbf{K} and the value \textbf{V}, and computes an attention matrix \( \mathbf{A} \) as:


\begin{equation}
\mathbf{A} = \text{softmax}\left(\frac{\mathbf{Q} \mathbf{K}^\top}{\sqrt{d_k}}\right)
\end{equation}
which is then multiplied by $\mathbf{V}$ to compute the output values.

To disrupt the attention of the model, we define a loss \( \mathcal{L}_{atn} \) that minimizes the alignment between the clean attention matrix \( \textbf{A}^{gt} \) and the adversarial attention matrix \( \textbf{A}^{adv} \) for a given layer $l$.
This loss is formulated as:

\begin{equation} 
\mathcal{L}_{atn} = \sum_{h=1}^{N_{h}} \frac{1}{(N_{t}-1)^2} \sum_{r=2}^{N_{t}} \sum_{c=2}^{N_{t}} A^{gt}_{h,r,c} \cdot A^{adv}_{h,r,c}
\end{equation}

where \( N_{h} \) is the number of attention heads, and \( N_{t} \) is the number of tokens. The terms \( A^{gt}_{h,r,c} \) and \( A^{adv}_{h,r,c} \) represent the attention values at head \( h \), row \( r \), and column \( c \) for the ground truth and adversarial attention matrices, respectively. The indices \( r, c \geq 2 \) indicate rows and columns of each attention map of shape \( N_t \times N_t \), where \( N_t \) corresponds to the number of visual tokens in each row/column, excluding the CLS token.
In our experiments, we attack the attention map from the last layer. In a set of preliminary experiments, we found that this strategy performs best, with earlier layers yielding slightly worse results.



By minimizing the dot product between the clean attention \( \textbf{A}^{gt} \) and the adversarial attention \( \textbf{A}^{adv} \), our loss function effectively alters adversarial attention making it orthogonal to clean attention. The main idea is that the original attention patterns are to be pushed away from the directions of high importance identified by the model.

This effect is applied to each token in the attention matrix. Specifically, for each token \( t \), minimizing the dot product between the attention row corresponding to \( t \) in \( \textbf{A}^{gt} \) and \( \textbf{A}^{adv} \) forces the adversarial perturbation to produce an orthogonal vector. As a consequence, where the clean attention would have a high value for a particular token, the adversarial perturbation induces a low value, thereby diminishing the attention on that token. Due to the nature of the softmax operation applied after the multiplication of query (\( \mathbf{Q} \)) and key (\( \mathbf{K} \)), this reduction in attention causes a redistribution of the attention scores across other tokens.

This mechanism is particularly effective because it not only changes the distribution of attention allocation but also directly impacts the magnitude of attention weights. By altering both aspects, the attack ensures that the attention is redistributed from the originally highly attended tokens to those that were previously less significant. This redistribution impairs the model ability to focus on relevant parts of the input, thereby degrading its overall performance and making the attack more powerful.

\subsection{Embedding Space Disruption}
To enhance the effectiveness of our adversarial attack, we not only manipulate the attention mechanism but also introduce a disruption in the embedding space. 
This approach allows us to directly alter the learned representations that are crucial for the model's performance across various tasks. By targeting the embedding space, we affect the foundational representation that informs subsequent decision-making layers.

Existing attacks in the embedding domain have access not just to the image but also to the text. Moreover, attacks are often carried out jointly on the two modalities, altering image and text \cite{zhang2022towards,lu2023set,he2023sa}.
CO-Attack \cite{zhang2022towards} crafts a multimodal embedding combining the two modalities and relies on the KL-Divergence to distance adversarial embeddings from unattacked image embeddings. SGA \cite{lu2023set} maximizes the misalignment, in terms of cosine distance, between text and image embedding. Finally, SA-Attack \cite{he2023sa} extends SGA through augmentation, thus leveraging input diversity to disrupt the alignment. 
In contrast, we attack a single modality, minimizing the dot product between adversarial embeddings and clean embeddings. Even without accessing or modifying the text embeddings, lowering the alignment of clean image embeddings and adversarial embeddings implicitly affects the alignment with textual embedding which are aligned by construction in models like CLIP~\cite{radford2021learning}.
We define the embedding loss, $\mathcal{L}_{emb}$, as the $L_{2}$ norm between the clean and adversarial embeddings:
\begin{equation}
\mathcal{L}_{emb} =  \left\| E^{gt} - E^{adv} \right\|_{2}  \end{equation}
To calculate the adversarial embedding $E^{adv}$, we first create an adversarial example by adding a perturbation $z^*$ to the input:
\begin{equation}
X^{adv} = X + z^*
\end{equation}

Then, we pass $X^{adv}$ through the model  to obtain $E^{adv}$:

\begin{equation}
E^{adv} = f_l(X^{adv}) = f_l(X + z^*)
\end{equation}

This gives us the adversarial embedding $E^{adv}$ from the transformer layer $l$ inside the vision encoder.
 

\subsection{Combined Loss Function}

To generate a robust adversarial perturbation, we combine the attention and embedding losses into a single objective:

\begin{equation}
\mathcal{L}_{comb} = \alpha \mathcal{L}_{atn} + \beta \mathcal{L}_{emb}
\end{equation}

where \( \alpha \) and \( \beta \) are hyperparameters that balance the contributions of each loss. By optimizing this combined loss, our approach effectively perturbs both the attention mechanism and embedding space, resulting in a powerful adversarial attack that compromises the integrity of foundational models and their downstream applications. The method optimizes the adversarial attack using Weighted Adam with an initial learning rate of $0.01$ for 250 iterations.
The losses $\mathcal{L}_{emb}$ and $\mathcal{L}_{atn}$ were balanced at every iteration by setting $\alpha=1; \beta = \alpha|\frac{\mathcal{L}_{atn}}{\mathcal{L}_{emb}}|$.
\begin{table}[!htb]
\centering
\small
\resizebox{.9\columnwidth}{!}{
\begin{tabular}{@{}lcccc@{}}
\toprule
\textbf{Attacks} & \textbf{ViT-B/16} & \textbf{ViT-B/32} & \textbf{ViT-L/14} & \textbf{Label} \\ \midrule
\textbf{FGSM}~\cite{goodfellow2014explaining} & 99.90 \% & 100\% & 99.70\% & \checkmark \\
\textbf{PGD}~\cite{pgd} & 99.90\% & 100\% & 99.80\% & \checkmark \\ \hline
\textbf{\EmbedAttack} & 97.90\% & 96.50\%& 98.10\%  & - \\
\textbf{\AttnAttackVisual} & 86.40\% & 85.50\%& 83.50\%  & - \\

\textbf{\CombAttack} & \textbf{98.40}\% & \textbf{97.40}\%& \textbf{98.30}\%  & - \\
\bottomrule
\end{tabular}
}
\caption{Attack success rate for classification on ImagenetV2 test set. $\epsilon = 2/255$}
\label{tab:classification}
\end{table}

\section{Experiments}
We evaluate our approach by covering a wide range of downstream tasks, assessing the impact of our attack.
We analyze the effects on classification, retrieval, captioning, zero-shot classification, depth estimation and semantic segmentation.
Note that, when different experiments share the same backbone (e.g., ViT-B/16), a single attack suffices to affect all of the described downstream tasks.

\paragraph{Classification}\label{sec:classification}
We evaluate our attack on image classification using Vision Transformers (ViTs) with different architectures. Specifically, we focus on ViT-B/16, ViT-B/32, and ViT-L/14 \cite{dosovitskiy2020image}, comparing the attack against two widely used targeted adversarial techniques: Fast Gradient Sign Method (FGSM)~\cite{goodfellow2014explaining} and Projected Gradient Descent (PGD)~\cite{pgd}. Tab.~\ref{tab:classification} shows that our attack yields slightly lower success rates. Nonetheless, differently from FGSM and PGD, our method does not require label information, highlighting its effectiveness and broader applicability.

\begin{table*}[t]
\centering
\begin{adjustbox}{width=0.98\textwidth}
\begin{tabular}{@{}lccc|ccc|ccc|ccc|ccc@{}}
\toprule
\textbf{Attacks} & \multicolumn{3}{c}{\textbf{COCO (Text Retrieval)}} & \multicolumn{3}{c}{\textbf{Flickr30K (Text Retrieval)}} & \multicolumn{3}{c}{\textbf{COCO (Image Retrieval)}} & \multicolumn{3}{c}{\textbf{Flickr30K (Image Retrieval)}} & \textbf{T} & \textbf{I} & \textbf{J} \\ 
& \textbf{R@1} & \textbf{R@5} & \textbf{R@10} 
& \textbf{R@1} & \textbf{R@5} & \textbf{R@10} 
& \textbf{R@1} & \textbf{R@5} & \textbf{R@10} 
& \textbf{R@1} & \textbf{R@5} & \textbf{R@10} 
& & & \\ \midrule
\textbf{Sep Attack}~\cite{zhang2022towards} & 69.10\% & 52.20\% & 43.64\% 
& 49.57\% & 23.99\% & 15.65\% 
& 78.64\% & 66.89\% & 60.66\% 
& 66.98\% & 51.83\% & 45.31\% 
& \checkmark & \checkmark & - \\ 
\textbf{SGA Attack}~\cite{lu2023set} & \underline{99.85}\% & \underline{99.42}\% & \underline{98.89}\% 
& \underline{99.14}\% & \underline{97.40}\% & \underline{95.73}\% 
& \underline{99.85}\% & \underline{99.37}\% & \underline{98.88}\% 
& \underline{99.10}\% & \underline{97.64}\% & \underline{96.16}\% 
& \checkmark & \checkmark & \checkmark \\

\textbf{SA Attack}~\cite{he2023sa} & 99.58\% & 98.74\% & 98.01\% 
& 98.65\% & 96.26\% & 93.09\% 
& 99.62\% & 99.06\% & 98.35\%
& 99.00\% & 96.87\% & 94.88\% 
& \checkmark & \checkmark & \checkmark \\
\textbf{Co-Attack}~\cite{zhang2022towards} & 97.98\% & 94.94\% & 93.00\% 
& 93.25\% & 84.88\% & 78.96\% 
& 98.80\% & 96.83\% & 95.33\% 
& 95.86\% & 90.83\% & 87.36\% 
& \checkmark & \checkmark & \checkmark \\
\hline
\textbf{Uni$_\text{A}$}\textsuperscript{*}~\cite{zhang2024universal} & 95.50\% & 91.42\% & 88.82\% 
& 91.90\% & 82.87\% & 78.66\% 
& 94.24\% & 90.11\% & 87.65\% 
& 91.14\% & 81.65\% & 79.96\% 
& - & \checkmark & - \\
\textbf{Mul$_\text{A}$}\textsuperscript{*}~\cite{zhang2024universal} & 95.50\% & 91.60\% & 88.87\% 
& 92.02\% & 82.04\% & 78.66\% 
& 96.13\% & 93.50\% & 91.76\% 
& 94.85\% & 90.42\% & 86.22\% 
& \checkmark & \checkmark & - \\
\textbf{ETU}\textsuperscript{*}~\cite{zhang2024universal} & 93.55\% & 89.01\% & 85.88\% 
& 88.47\% & 78.61\% & 73.07\% 
& 94.25\% & 91.30\% & 89.31\% 
& 92.69\% & 87.50\% & 84.89\% 
& \checkmark & \checkmark & - \\ 
\textbf{SGA Attack}~\cite{lu2023set} & 97.30\% & 94.02\% & 91.17\% 
& 94.11\% & 88.89\% &83.64\% & 97.72\% & 94.65\% & 91.86\% & 95.91\% & 90.10\% & 85.98\%
& - & \checkmark & - \\
\textbf{Co-Attack}~\cite{zhang2022towards}  & 94.93\% & 90.19\% & 86.52\% 
& 87.73\% & 78.09\% & 72.05\% & 95.88\% & 91.58\% & 88.55\%  & 91.72\% & 83.32\% & 78.67\%
& - & \checkmark & - \\
\textbf{\EmbedAttack} & \textbf{99.81}\% & 99.50\% & \textbf{99.29}\%
& \textbf{99.26}\% & 98.03\% & 97.36\%
& 99.42\% & \textbf{98.72}\% & \textbf{98.21}\%
& 98.61\% & 97.01\% & 95.81\% & - & \checkmark & - \\ 
\textbf{\AttnAttackVisual} & 97.10\% & 93.54\% & 91.33\%
& 89.82\% & 80.48\% & 74.80\%
& 95.36\% & 90.63\% & 87.26\%
& 89.79\% & 78.37\% & 72.20\% & - & \checkmark & - \\ 
\textbf{\CombAttack}  & \textbf{99.81}\% & \textbf{99.53}\% & \textbf{99.29}\%
& 99.14\% & \textbf{98.55}\% & \textbf{97.66}\%
& \textbf{99.43}\% & 98.61\% & 98.11\%
& \textbf{98.87}\% & \textbf{97.41}\% & \textbf{96.16}\% & - & \checkmark & - \\ 

\end{tabular}
\end{adjustbox}
\caption{Attacks to CLIP on retrieval on COCO and Flickr30K. Universal perturbation attack with a 12/255 budget are marked with\textsuperscript{*}. Remaining methods use $\epsilon = 2/255$. Best results not using joint information are in \textbf{bold}.  Best results using joint information are \underline{underlined}.\label{tab:combined_results}}
\end{table*}

\paragraph{Image/Text Retrieval}\label{sec:retrieval}

In Tab.~\ref{tab:combined_results}, we present a comparison of our method with state-of-the-art approaches on the image/text retrieval tasks using the COCO \cite{lin2015microsoft} and Flickr30K \cite{young2014image} datasets by attacking CLIP (ViT-B/16). As evaluation metric, we use attack success rate at \textbf{R@1}, \textbf{R@5}, and \textbf{R@10}. The term \textbf{TR-R@K} refers to the percentage of adversarial images for which the correct caption (among the ground truth captions) is not retrieved in the top-K by the VLP model. Similarly, \textbf{IR-R@K} indicates the percentage of texts for which the correct image is not found in the top-K results when the attack is performed.
To enable this attack with our architecture, we attack all images in the gallery, leaving the input text unaltered. 
Columns labeled ``T'', ``I" and ``J" indicate whether the attack targets text (T), image (I), or both jointly (J).
For both the SGA attack \cite{lu2023set} and Co-Attack \cite{zhang2022towards} baselines, results are provided for two distinct settings: the image-only attack and the joint image-and-text attack. The image-only setting applies perturbations solely to the image (I), while the joint setting perturbs both the image and text modalities (I, T).

In the text retrieval task, our proposed method achieves a high attack success rate, surpassing joint modality methods such as SGA Attack \cite{lu2023set}, SA Attack \cite{he2023sa}, and Co-Attack \cite{zhang2022towards}.
Furthermore, when SGA and Co-Attack operate in the image-only setting, their performance drops significantly. Our method consistently outperforms them, with the only exception being that SGA achieves a slightly higher attack success rate at R@1 on COCO.
Similarly, for image retrieval, our method performs on par with the joint modality competitors, achieving high attack success rates, with R@1 values of 99.43\% on the COCO dataset and 98.87\% on the Flickr30K dataset.
Again, in the image-only setting, our method outperforms both SGA and Co-Attack.

\begin{figure*}[t] 
    \centering
    \tiny
    \begin{tabular}{*{6}{>{\centering\arraybackslash}m{0.14\textwidth}}} 
         & \includegraphics[width=\linewidth, height= 0.8\linewidth]{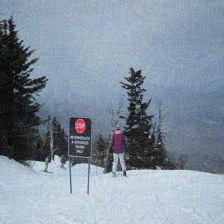} &
        \includegraphics[width=\linewidth, height= 0.8\linewidth]{images/caption/50.png} &
        \includegraphics[width=\linewidth, height= 0.8\linewidth]{images/caption/5.png} &
        \includegraphics[width=\linewidth, height= 0.8\linewidth]{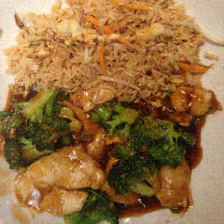} &
        \includegraphics[width=\linewidth, height= 0.8\linewidth]{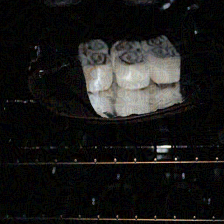} \\ \hline
        \scriptsize Predicted caption (unattacked) & 
        \scriptsize A person on skis standing on a snow covered slope & 
        \scriptsize two girls holding donuts up to their eyes & 
        \scriptsize A pizza with arugula and prosciutto on top &
        \scriptsize A plate of chicken, broccoli and rice & 
        \scriptsize A tray of cinnamon rolls in an oven \\
        \hline
        \scriptsize \AttnAttackVisual & \scriptsize a photograph of a glass of ice with a red light in it & 
        \scriptsize A picture of a person in a pool of water & 
        \scriptsize A bunch of grapes in a glass of water & 
        \scriptsize A painting of a woman with a flower in her hair & 
        \scriptsize A black and white image of a black and white image of a black and white image of a \\
        \hline
        \scriptsize SGA Attack~\cite{lu2023set} & \scriptsize A group of people on skis on a snow covered slope & 
        \scriptsize A man and a woman drinking a donut & 
        \scriptsize A plate of food on a wooden cutting board & 
        \scriptsize A plate of food on a table & 
        \scriptsize A piece of bread in an oven \\
        \hline
        \scriptsize Co-Attack~\cite{zhang2022towards} & \scriptsize A person on skis on a snowy slope & 
        \scriptsize A girl wearing a pink shirt & 
        \scriptsize A sandwich on a plate & 
        \scriptsize A plate of food on a table & 
        \scriptsize a black and white picture of a tooth brush in a dark room \\
        \hline
    \end{tabular}
    \caption{Comparison of adversarial and original COCO images with predicted and adversarial captions from BLIP-2. Adversarial examples were crafted with $\epsilon = 8/255$. The first row shows the predicted captions, the remaining rows show captions generated by attacking with: Our Method (second row), SGA (third row) and Co-attack (fourth row).}
    \label{fig:comparison}
\end{figure*}

\paragraph{Image Captioning}\label{sec:captioning}
Table~\ref{tab:captioning} presents the impact of adversarial attacks on image captioning for the BLIP-2 model\footnote{We used the “BLIP-2-t5/pretrain-flant5xl-vitL” variant.}~\cite{li2023blip}, evaluated on COCO and Flickr30k. Performance is measured using BLEU (1–4), METEOR, ROUGE, CIDEr, and SPICE, which collectively measure the quality and accuracy of the generated captions compared to reference captions.
Our attack causes a severe performance drop across all metrics. On COCO, BLEU-2 falls from 0.6212 to 0.2004, signaling a collapse in bigram precision, while CIDEr, which quantifies similarity to human captions, drops from 1.2912 to 0.0716. The degradation extends to Flickr30k, further demonstrating the vulnerability of captioning models, especially those relying on foundation models like CLIP, under adversarial attacks.
A key distinction between our method and existing adversarial approaches lies in what is being attacked. 
Even though state-of-art methods, such as SGA \cite{lu2023set} and Co-Attack \cite{zhang2022towards}, optimize adversarial perturbations by jointly targeting both the text and image modalities, their focus is exclusively on perturbing the embedding space. In contrast, we attack both embeddings and the attention mechanism simultaneously. This dual disruption results in a more severe degradation of model performance.
In Fig.~\ref{fig:comparison}, we show the effects of adversarial perturbations for the captioning task using COCO and the BLIP-2 model. The first row displays adversarial examples generated with a perturbation magnitude of $\epsilon = 8/255$. These adversarial examples are visually indistinguishable from the original images but induce significant errors in the model’s predictions.
The second row shows the original captions generated by BLIP-2 for clean images, where the model accurately describes the content. In contrast, the third row presents the adversarial captions generated after applying the attention-based perturbations, which lead the model to produce inaccurate and unrelated descriptions.
We also report captions generated using SGA Attack and CO-Attack.

\begin{table}[t]
\centering

\resizebox{\columnwidth}{!}{
\begin{tabular}{lcccccccc}
\toprule
\textbf{Metric} & \textbf{BLEU-1} & \textbf{BLEU-2} & \textbf{BLEU-3} & \textbf{BLEU-4} & \textbf{METEOR} & \textbf{ROUGE} & \textbf{CIDER} & \textbf{SPICE} \\
\midrule
\multicolumn{9}{c}{\textbf{COCO Dataset}} \\
\hline
\textbf{Clean} & 0.7595 & 0.6212 & 0.4877 & 0.3725 & 0.2819 & 0.5888 & 1.2912 & 0.2255 \\
\textbf{Co-Attack}~\cite{zhang2022towards} & 0.6654 & 0.5220 & 0.3925 & 0.2895 & 0.2380 & 0.5264 & 0.9810 & 0.1790\\
\textbf{SGA}~\cite{lu2023set} & 0.6635 & 0.5012 & 0.3676 & 0.2653 & 0.2244 & 0.5071 & 0.8806 & 0.1644\\
\textbf{\AttnAttackVisual} & 0.4022 & 0.2025 & 0.1005 & 0.0600 & 0.0938 & 0.3039 & 0.0788 & 0.0261\\
\textbf{\EmbedAttack} & 0.4815 & 0.2766 & 0.1571 & 0.0934 & 0.1272 & 0.3527 & 0.1936 & 0.0590\\
\textbf{\CombAttack} & \textbf{0.4000} & \textbf{0.2004} & \textbf{0.0983} & \textbf{0.0558} & \textbf{0.0878} & \textbf{0.2949} & \textbf{0.0716} & \textbf{0.0252}\\

\midrule

\multicolumn{9}{c}{\textbf{FLICKR Dataset}} \\
\hline
\textbf{Clean} & 0.6997 & 0.5392 & 0.3970 & 0.2874 & 0.2179 & 0.5058 & 0.6741 & 0.1567 \\
\textbf{Co-Attack}~\cite{zhang2022towards} & 0.6499 & 0.4693 & 0.3295 & 0.2287 & 0.1903 & 0.4563 & 0.5198 & 0.1330\\
\textbf{SGA}~\cite{lu2023set} & 0.6169 & 0.4266 & 0.2900 & 0.1948 & 0.1723 & 0.4308 & 0.4180 & 0.1127\\
\textbf{\AttnAttackVisual} & 0.3846 & 0.1770 & 0.0757 & 0.0388 & 0.0713 & 0.2632 & 0.0350 & 0.0197\\
\textbf{\EmbedAttack} & 0.4892 & 0.2889 & 0.1765 & 0.1135 & 0.1173 & 0.3385 & 0.1361 & 0.0583\\
\textbf{\CombAttack} & \textbf{0.03598} & \textbf{0.1640} & \textbf{0.0705} & \textbf{0.0359} & \textbf{0.0662} & \textbf{0.2463} & \textbf{0.0332} & \textbf{0.0183}\\

\bottomrule
\end{tabular}}
\caption{Impact of adversarial attacks on image captioning metrics for the BLIP-2 model\tablefootnote{https://huggingface.co/Salesforce/blip2-flan-t5-xl} with $\epsilon = 8/255$. Evaluated on 1,000 random images from COCO (val) and FLICKR datasets. Adversarially perturbed images are generated using the CLIP-ViT-L model.}
\label{tab:captioning}
\end{table}


\paragraph{Zero-Shot classification}\label{sec:zeroshot}
In Tab.~\ref{tab:zero_shot_1}, we evaluate the performance of different adversarial attack methods on zero-shot classification tasks.
Following prior work~\cite{wang2024benchmarking}, we test the zero-shot capability of CLIP with different visual backbones, namely ViT-B/16 and ViT-B/32. In the tables, we report the classification accuracy after the attack.
The evaluations are performed on 1,000 random images from the ImageNet 2012 \cite{ILSVRC15} dataset, with an adversarial perturbation budget of $\epsilon = 8/255$.
Notably, our proposed approach effectively attacks all images, yielding a 0\% classification accuracy, as well as DeepFool, BIM and MIM. The only white-box attack that is not able to achieve a perfect misclassification is FGSM (6.7\% and 10.2\% on ViT-B/16 and ViT-B/32). Black-box attacks instead are less effective as they do not manage to lower the accuracy below 40\%.
It has to be noted that all white-box competitors have access to both images and text during the attack, contrary to our method that only attacks images, without any knowledge about text. We specify whether a model has access to text (\textbf{T}) and/or images (\textbf{I}) in Tab.~\ref{tab:zero_shot_1}.


\begin{table}
    \centering
    \small
    \resizebox{.85\columnwidth}{!}{
    \begin{tabular}{lcccc}
    \toprule
        \textbf{Attacks} & ~~~~~\textbf{ViT-B/16}~~~~~ & ~~~~~\textbf{ViT-B/32}~~~~~ & \textbf{T} & \textbf{I}\\ \hline
        \textbf{NES}*~\cite{ilyas2018black} & 42.80\% & 41.00\% & - & \checkmark \\ 
        \textbf{SPSA}*~\cite{uesato2018adversarial} & 43.50\% & 40.50\% & - & \checkmark \\ 
        \textbf{FGSM}~\cite{goodfellow2014explaining} & 6.70\% & 10.20\% & \checkmark & \checkmark  \\ 
        \textbf{DeepFool}~\cite{moosavi2016deepfool} & \textbf{0\%} & \textbf{0\%} & \checkmark & \checkmark  \\ 
        \textbf{BIM}~\cite{kurakin2018adversarial} & \textbf{0\%} & \textbf{0\%} &  \checkmark & \checkmark  \\ 
        \textbf{MIM}~\cite{dong2018boosting} & \textbf{0\%} & \textbf{0\%} & \checkmark & \checkmark  \\
        \textbf{\AttnAttackVisual} & 1.00\% & 1.20\% & - & \checkmark  \\
        \textbf{\EmbedAttack} & 0.10\% & 0.10\% & - & \checkmark  \\
        \textbf{\CombAttack} & \textbf{0\%} & \textbf{0\%} & - & \checkmark  \\
      
        \bottomrule  
        
    \end{tabular}
    }
    \caption{Zero shot accuracy on ImageNet using $\epsilon = 8/255$. Black box attacks denoted with ${}^*$. }
    \label{tab:zero_shot_1}
\end{table}

\begin{figure}[t] 
    \centering
    \tiny
\resizebox{\columnwidth}{!}{
    \begin{tabular}{*{5}{>{\centering\arraybackslash}m{0.18\textwidth}}} 
        \includegraphics[width=\linewidth, height= 0.8\linewidth]{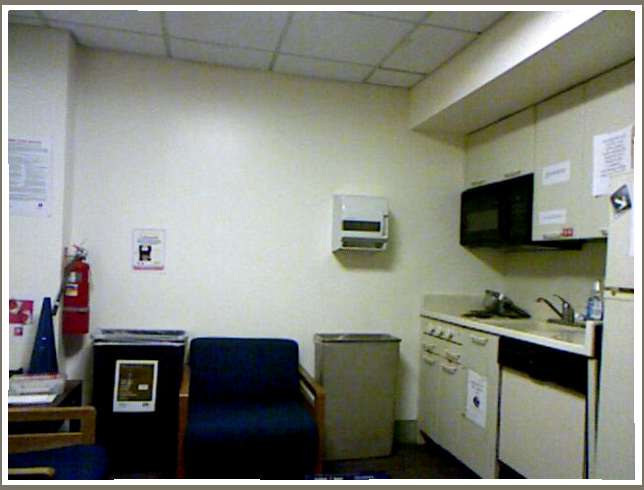} &
        \includegraphics[width=\linewidth, height= 0.8\linewidth]{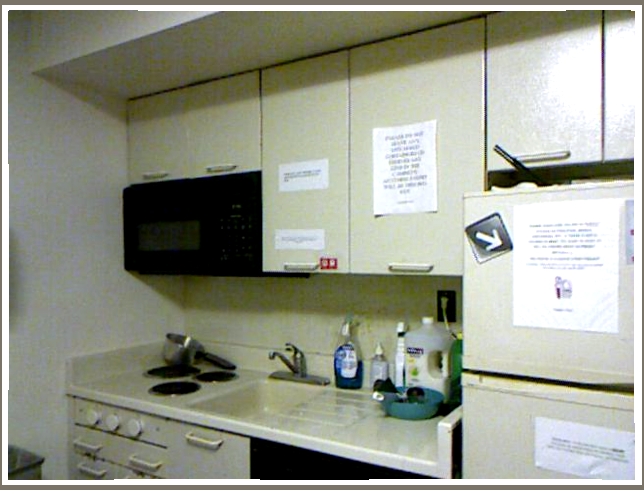} &
        \includegraphics[width=\linewidth, height= 0.8\linewidth]{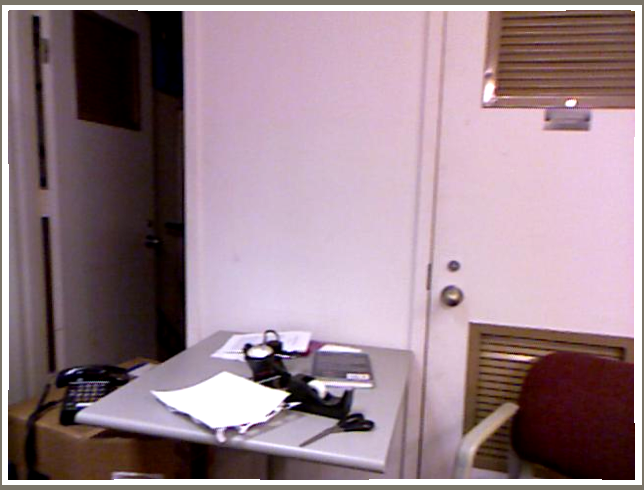} &
        \includegraphics[width=\linewidth, height= 0.8\linewidth]{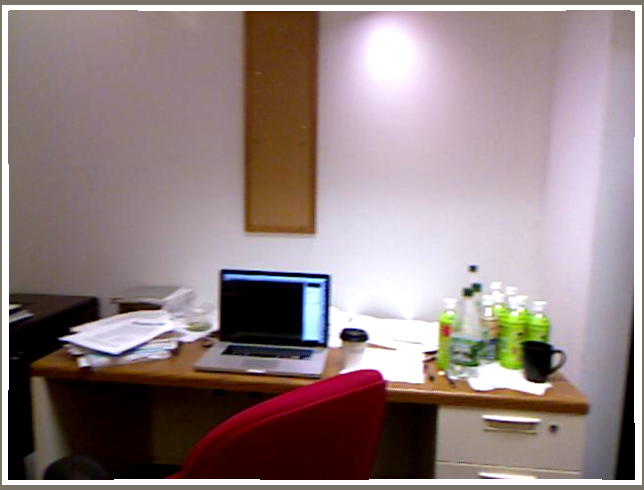} &
        \includegraphics[width=\linewidth, height= 0.8\linewidth]{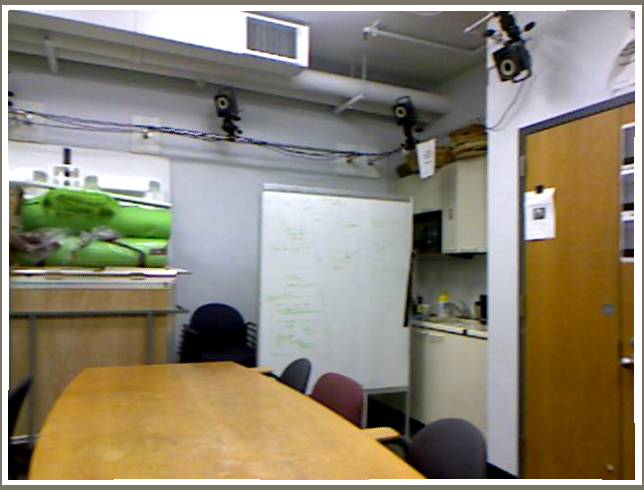} \\

        \includegraphics[width=\linewidth, height= 0.8\linewidth]{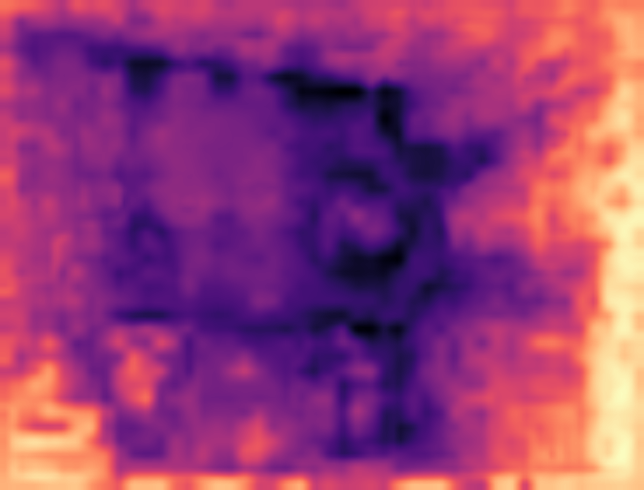} &
        \includegraphics[width=\linewidth, height= 0.8\linewidth]{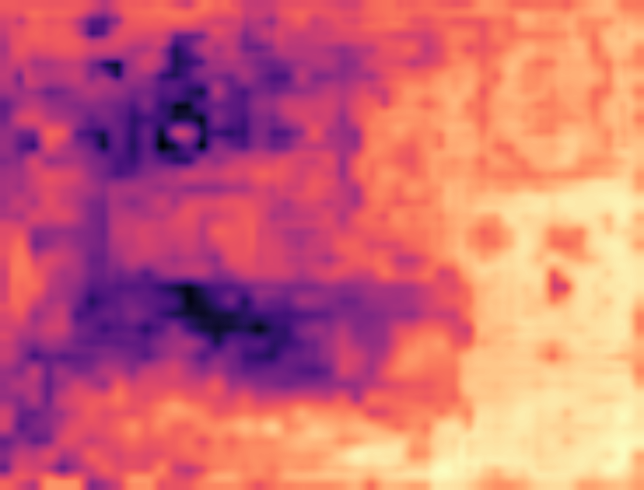} &
        \includegraphics[width=\linewidth, height= 0.8\linewidth]{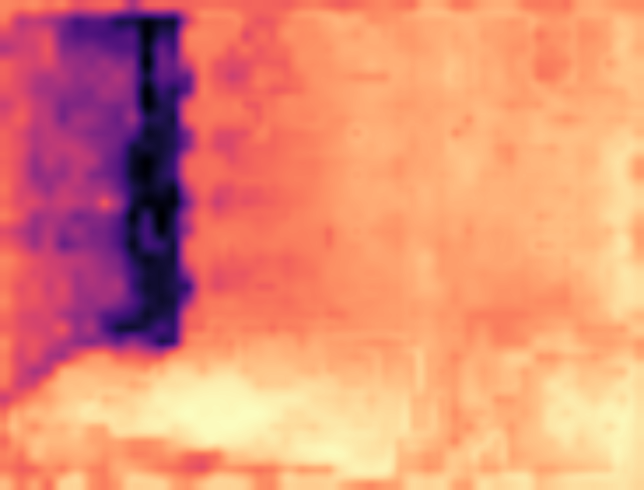} &
        \includegraphics[width=\linewidth, height= 0.8\linewidth]{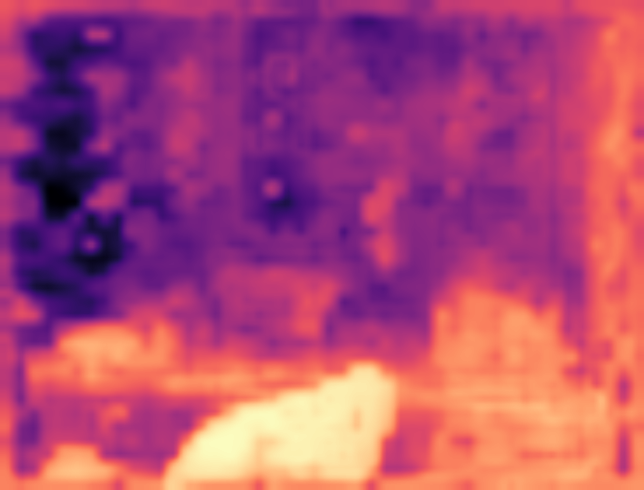} &
        \includegraphics[width=\linewidth, height= 0.8\linewidth]{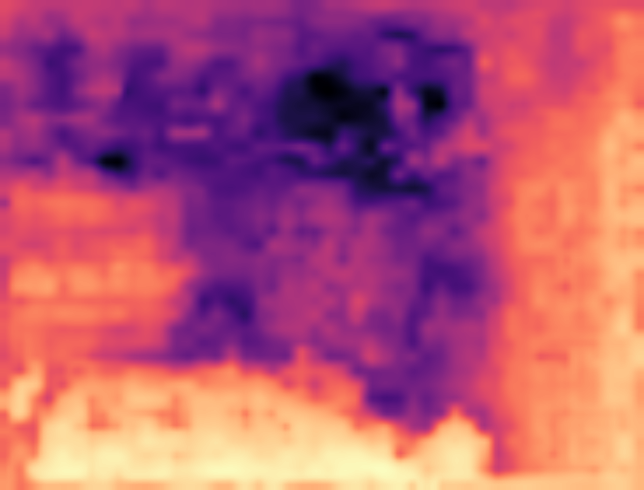} \\
        \includegraphics[width=\linewidth, height= 0.8\linewidth]{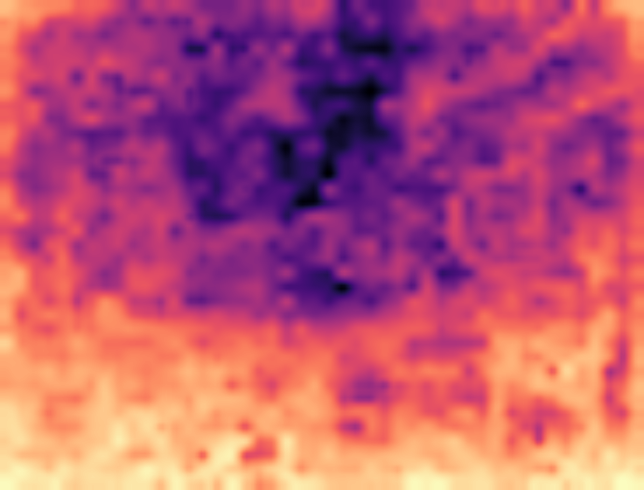} &
        \includegraphics[width=\linewidth, height= 0.8\linewidth]{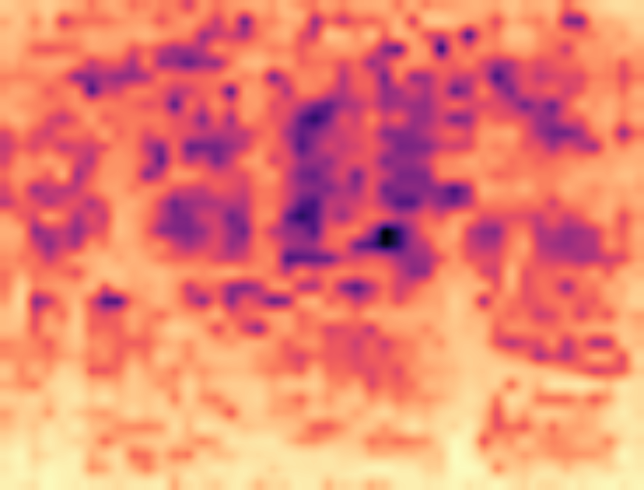} &
        \includegraphics[width=\linewidth, height= 0.8\linewidth]{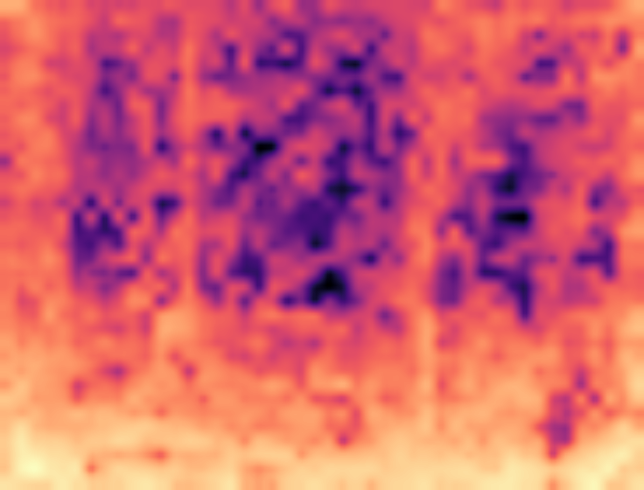} &
        \includegraphics[width=\linewidth, height= 0.8\linewidth]{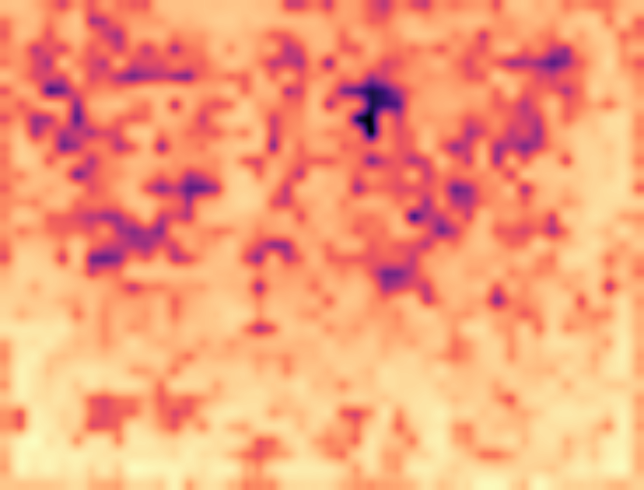} &
        \includegraphics[width=\linewidth, height= 0.8\linewidth]{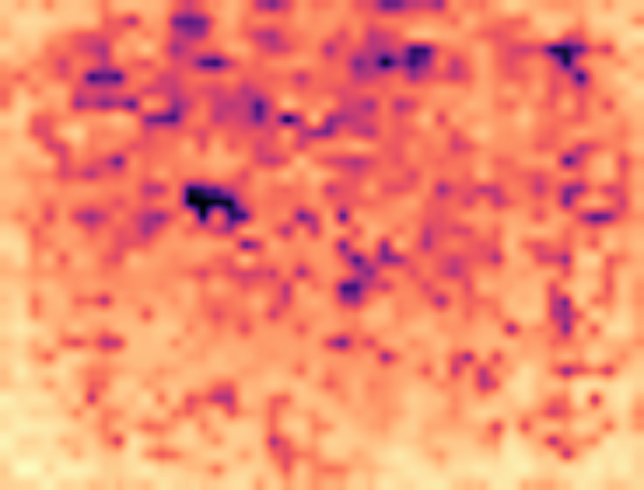} \\

    \end{tabular}
    }
    \caption{Estimated depths with unattacked (second row) and attacked (third row) images using DinoV2 ViT-S/14 model.}
    \label{fig:depth_qualitative}
\end{figure}

\paragraph{Depth Estimation}\label{sec:depth}
For depth estimation, we investigate the impact of the proposed attack on models\footnote{ViT-S/14 and ViT-B/14 (1 layer)} that use DINOv2. In Tab.~\ref{tab:depth_estimation}, we present the results on the NYU-Depth v2 dataset \cite{silberman2012indoor} for our method compared to the task-specific attack DepthPGD~\cite{kowalczuk2024benchmarking}, which attacks the model's head pixel-wise.
We report the post-attack Root Mean Squared Error (RMSE), which significantly increases compared to the one of the unattacked models.
Although our approach does not outperform DepthPGD, it is agnostic to the model architecture and, as shown in Fig.~\ref{fig:depth_qualitative}, it introduces significant distortions.
It is interesting to notice that our attention-based attack obtains the best results compared to our other proposed variations. In fact, the attack that targets only the embedding does not prove to be as effective, even in combination with the attention loss. We impute this to the fact that depth estimation is a very fine-grained task, where rather than compressing information into a latent vector (as in, say, classification), the purpose is to capture spatial relations among pixels. Attacking attention hinders the capabilities of the model to perform such reasoning.

\begin{table}[t]
\tiny
\centering
\resizebox{.85\columnwidth}{!}{
\begin{tabular}{lccc}
\toprule
\textbf{Attacks} & \textbf{ViT-S/14} & \textbf{ViT-B/14} & \textbf{Task-agnostic} \\ \midrule
\textit{No attack} & 0.49 & 0.46 & -\\
\textbf{\DepthPGD}~\cite{kowalczuk2024benchmarking} & \textbf{2.60} & \textbf{2.74} & - \\
\textbf{\AttnAttackVisual} & 1.50 & 1.35 & \checkmark \\
\textbf{\EmbedAttack} & 1.44 & 1.07 & \checkmark \\
\textbf{\CombAttack} & 1.46 & 1.34 & \checkmark \\

\bottomrule
\end{tabular}
}
\caption{RMSE for depth estimation for different attacks on the NYU-Depth v2 dataset. Attack budget $\epsilon=8/255$.}
\label{tab:depth_estimation}
\end{table}

\begin{figure}[t] 
    \centering
    \tiny
\resizebox{\columnwidth}{!}{
    \begin{tabular}{*{5}{>{\centering\arraybackslash}m{0.18\textwidth}}} 
        \includegraphics[width=\linewidth, height= 0.8\linewidth]{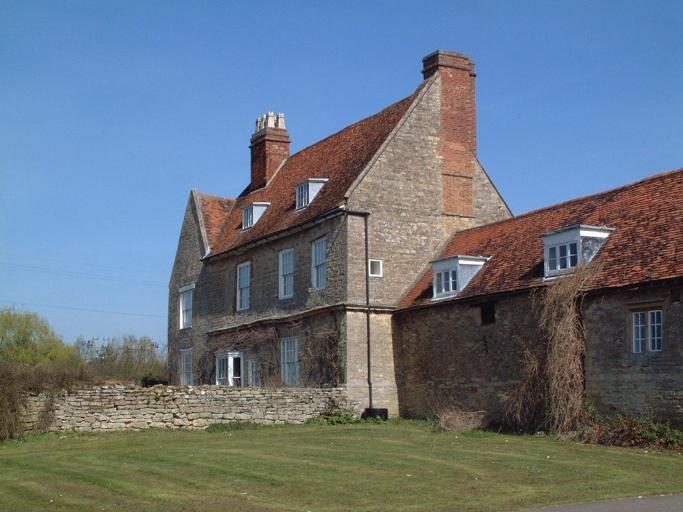} &
        \includegraphics[width=\linewidth, height= 0.8\linewidth]{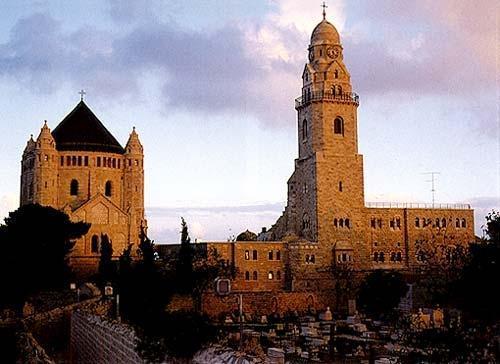} &
        \includegraphics[width=\linewidth, height= 0.8\linewidth]{images/segmentation/ADE_val_00000003.jpg} &
        \includegraphics[width=\linewidth, height= 0.8\linewidth]{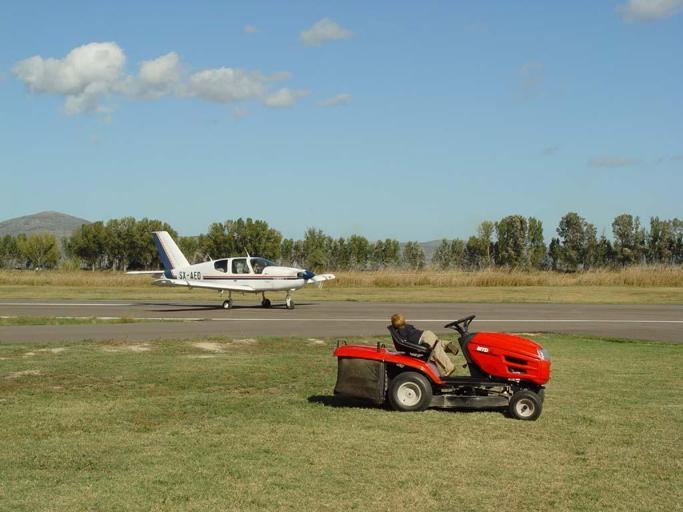} &
        \includegraphics[width=\linewidth, height= 0.8\linewidth]{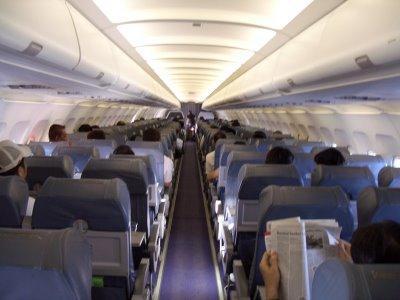} \\

       \includegraphics[width=\linewidth, height= 0.8\linewidth]{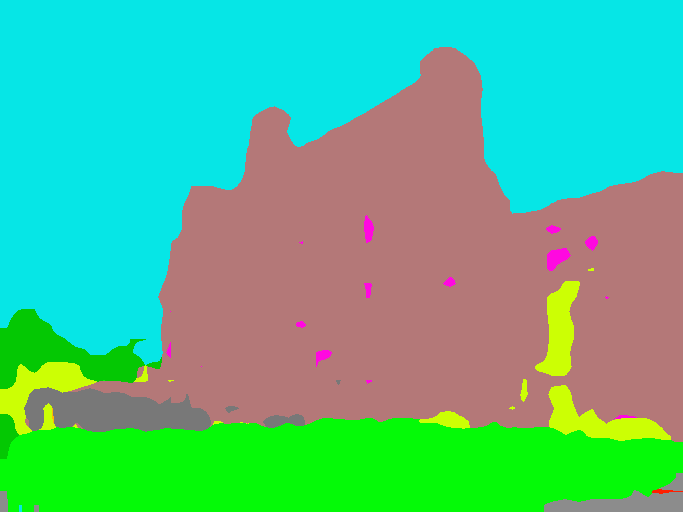} &
        \includegraphics[width=\linewidth, height= 0.8\linewidth]{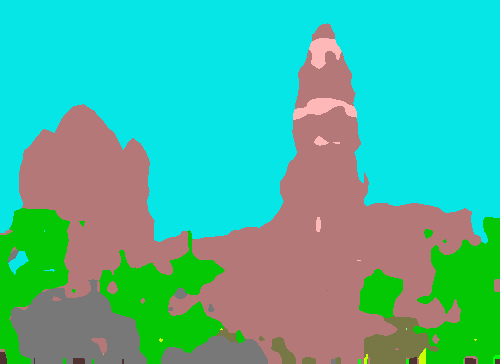} &
        \includegraphics[width=\linewidth, height= 0.8\linewidth]{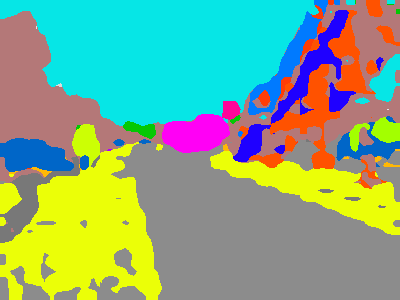} &
        \includegraphics[width=\linewidth, height= 0.8\linewidth]{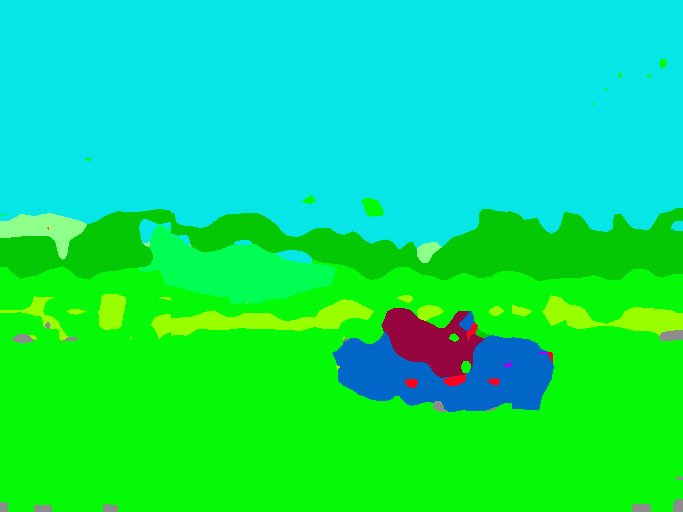} &
        \includegraphics[width=\linewidth, height= 0.8\linewidth]{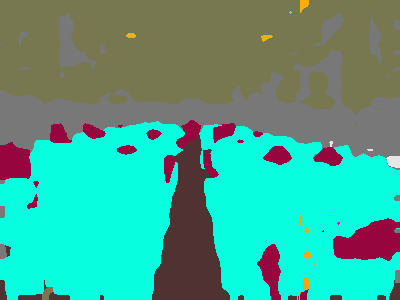} \\
        \includegraphics[width=\linewidth, height= 0.8\linewidth]{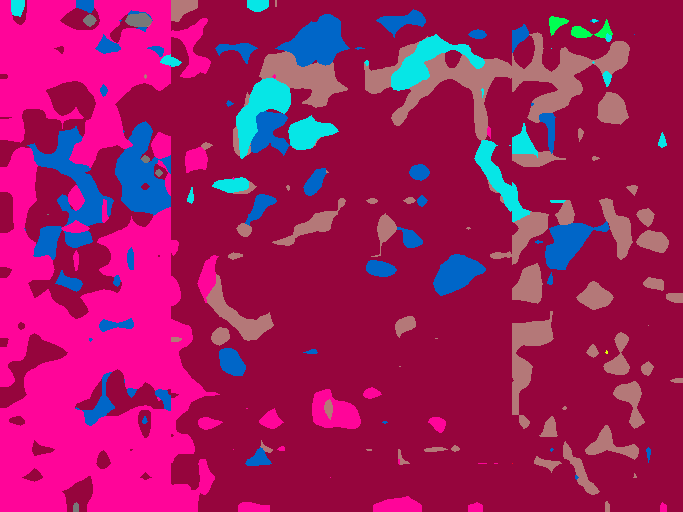} &
        \includegraphics[width=\linewidth, height= 0.8\linewidth]{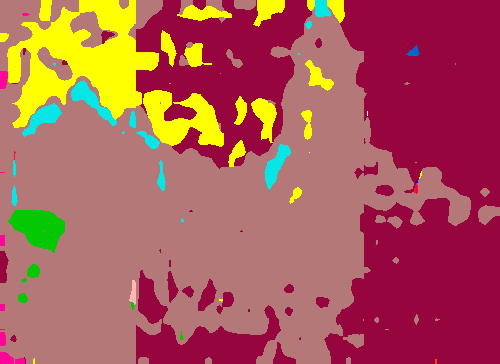} &
        \includegraphics[width=\linewidth, height= 0.8\linewidth]{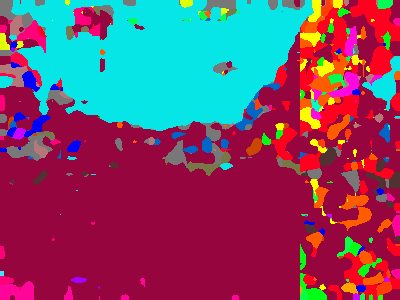} &
        \includegraphics[width=\linewidth, height= 0.8\linewidth]{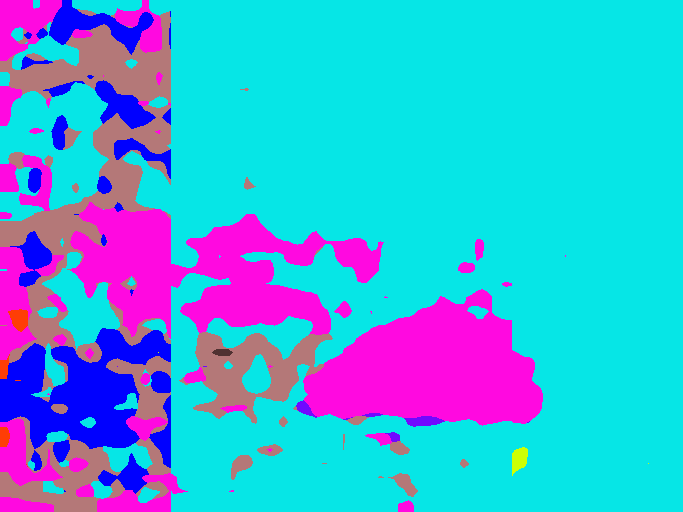} &
        \includegraphics[width=\linewidth, height= 0.8\linewidth]{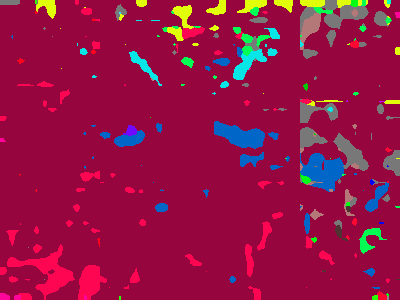} \\
    \end{tabular}
    }
    \caption{Semantic segmentations with unattacked (second row) and attacked (third row) images with the DinoV2 ViT-S/14 model.}
    \label{fig:segmentation_qualitative}
\end{figure}

\paragraph{Semantic Segmentation}\label{sec:segmentation}
Similarly to depth estimation, for semantic segmentation, we use ViT models\footnote{ViT-S/14 and ViT-B/14 (linear)} that utilize DINOv2. In Tab.\ref{tab:segmentation}, we compare our method against SegPGD~\cite{kowalczuk2024benchmarking}, a task-specific attack designed to disrupt segmentation performance by directly targeting the model’s output. We evaluate on the ADE20K dataset~\cite{zhou2019semantic}, measuring performance in terms of mean Intersection over Union (mIoU). Although SegPGD achieves near-complete degradation, reducing mIoU to 0.01, our approach is able to slightly improve, lowering it to 0.008.
Again, attention-based attacks prove to be more effective than embedding-based ones, confirming the trend of Tab.~\ref{tab:depth_estimation}.
In Fig.~\ref{fig:segmentation_qualitative}, we show qualitative results of semantic segmentation by the DinoV2 model for clean and attacked images.


\begin{table}[t]
\tiny
\centering
\resizebox{.85\columnwidth}{!}{
\begin{tabular}{lccc}
\toprule
\textbf{Attacks} & \textbf{ViT-S/14} & \textbf{ViT-B/14} & \textbf{Task-agnostic}\\ \midrule
\textit{No attack} & 0.420 & 0.450 & - \\
\textbf{\SegPGD}~\cite{kowalczuk2024benchmarking} & 0.010 & 0.010 & - \\
\textbf{\AttnAttackVisual} & \textbf{0.008} & 0.010 & \checkmark \\
\textbf{\EmbedAttack} & 0.026 & 0.044 & \checkmark \\
\textbf{\CombAttack} & \textbf{0.008} & \textbf{0.009} & \checkmark \\

\bottomrule
\end{tabular}
}
\caption{mIoU for semantic segmentation for different attacks on ADE20K. Attack budget $\epsilon=8/255$.}
\label{tab:segmentation}
\end{table}

\paragraph{Attacking Attention vs. Attacking Embeddings}
From the experiments, it emerges that different downstream tasks are affected in different ways depending on which parts of the attention layers we target. It is interesting to notice that "semantic" tasks, such as classification (Tab.~\ref{tab:classification}) and cross-modal retrieval (Tab.~\ref{tab:combined_results}) are more easily disrupted by attacking embeddings rather than attention. This is understandable, as token-to-token relationships are not so important for the task as instead the overall representation of the image is.
On the contrary, more fine-grained tasks such as captioning (Tab.~\ref{tab:captioning}), depth estimation (Tab.~\ref{tab:depth_estimation}) and segmentation (Tab.~\ref{tab:segmentation}), suffer more when attacked using attention since spatial relations between objects or pixels have a direct effect on the outputs.
Using a combination of both results to be effective throughout most downstream tasks.

\paragraph{Transferability}\label{sec:transferability}
Adversarial transferability is a crucial yet challenging aspect of adversarial robustness, particularly important in real-world scenarios. In Tab.~\ref{tab:transferability}, we compare our method against existing adversarial attacks, including SGA~\cite{lu2023set} and Co-Attack~\cite{zhang2022towards}, for image-to-text and text-to-image retrieval tasks on Flickr30k.
We use as source model CLIP with ViT-B/16 backbone and as target model ALBEF, TCL and CLIP with a CNN-based backbone (ResNet101).
While SGA achieves the highest transfer success rates, it is important to note that this advantage largely stems from its joint optimization with text, allowing it to perturb both modalities. However, the absolute retrieval degradation remains relatively low, limiting its threat level in real-world scenarios. This suggests that adversarial transferability remains an open problem requiring further investigation.

\begin{table}[t]
\resizebox{\columnwidth}{!}{
\begin{tabular}{l|l|ccc|ccc}
    \toprule

        \multirow{2}{*}{\textbf{\fontsize{10pt}{\baselineskip}\selectfont{Target}}} & \multirow{2}{*}{\textbf{\fontsize{10pt}{\baselineskip}\selectfont{Method}}} & \multicolumn{3}{c|}{\textbf{\fontsize{10pt}{\baselineskip}\selectfont{Image-to-Text}}} & \multicolumn{3}{c}{\textbf{\fontsize{10pt}{\baselineskip}\selectfont{Text-to-Image}}}   \\
         &   & R@1  & R@5  & R@10    & R@1  & R@5  & R@10 \\
      \hline
      \multirow{6}{*}{CLIP$_{\rm ViT}$ $\rightarrow$ ALBEF} 
            & Sep-Attack  & 2.50  & 0.40  & 0.10 & 4.93 & 1.44 & 1.01 \\
        & Co-Attack  & 2.50 & 0.60 & 0.20 & 5.80 & 1.78 & 1.11 \\
      & SGA & \textbf{3.86} & \textbf{0.70} & \textbf{0.30} & \textbf{7.69} & \textbf{2.73} & \textbf{1.52}  \\
      & \AttnAttackVisual & 2.61 & 0.20 & 0.20 & 5.29 & 1.29 & 0.99  \\
      & \EmbedAttack & 2.92 & 0.40 & 0.20 & 5.05 & 1.54 & 0.99  \\
      & \CombAttack & 2.40 & 0.40 & 0.10 & 5.10 & 1.54 & 0.97  \\
      \hline 
      \multirow{6}{*}{CLIP$_{\rm ViT}$ $\rightarrow$ TCL} 
      & Sep-Attack  & 4.85  &  0.20 & \textbf{0.20} & 8.17 & 2.27 & 1.46 \\
      & Co-Attack  & 5.27 & 0.40 & \textbf{0.20} & 9.12 & 2.75 & 1.48  \\
      & SGA & \textbf{6.43} & 0.60 & \textbf{0.20} & \textbf{10.93} & \textbf{3.47} & \textbf{2.05}   \\

      & \AttnAttackVisual & 5.42 & 0.60 & \textbf{0.20} & 10.55 & 2.93 & 1.78  \\  
      & \EmbedAttack & 5.63 & \textbf{0.70} & 0.10 & 10.59 & 2.91 & 1.76  \\
      & \CombAttack & 5.53 & \textbf{0.70} & 0.10 & 10.76 & 2.93 & 1.66  \\

      \hline
      \multirow{6}{*}{CLIP$_{\rm ViT}$ $\rightarrow$ CLIP$_{\rm CNN}$}    & Sep-Attack  & 5.36  & 1.16  & 0.72 & 8.44 & 2.35 & 1.54 \\
      & Co-Attack  & 7.66 & 1.90 & 1.44 & 9.37 & 3.90 & 2.53   \\
      & SGA & \textbf{11.24} & \textbf{5.39} & \textbf{2.68} & \textbf{15.68} & \textbf{6.88} & \textbf{5.08}   \\    
     
      & \AttnAttackVisual & 6.13 & 1.48 & 1.24 & 8.30 & 3.30 & 2.12  \\
      & \EmbedAttack & 6.64 & 1.59 & 0.82 & 8.75 & 3.06 & 1.83  \\
      & \CombAttack & 6.39 & 1.80 & 0.93 & 8.58 & 2.86 & 2.30  \\

    \bottomrule
        
    \end{tabular}
}
    \caption{Transferability experiment on Flicr30K dataset.}
    \label{tab:transferability}
\end{table}

\section{Conclusions}
We presented a novel task-agnostic adversarial attack method tailored for attention-based vision foundation models. Our approach targets both the attention mechanism and the embedding space within these models, resulting in significant disruptions of performance across a variety of downstream tasks.
Our experiments, conducted on widely-used datasets show that our method achieves high attack success rates using only image perturbations, without the need for textual data or joint multi-modal optimization. 
The simplicity and effectiveness of the proposed methods highlights potential issues of relying on widespread foundation models.

\section*{Acknowledgements}
This work was partially supported by project “Collaborative Explainable
neuro-symbolic AI for Decision Support Assistant, CAI4DSA, CUP B13C23005640006.”

{
    \small
    \bibliographystyle{ieeenat_fullname}
    \bibliography{main}

\begin{thebibliography}{51}
\providecommand{\natexlab}[1]{#1}
\providecommand{\url}[1]{\texttt{#1}}
\expandafter\ifx\csname urlstyle\endcsname\relax
  \providecommand{\doi}[1]{doi: #1}\else
  \providecommand{\doi}{doi: \begingroup \urlstyle{rm}\Url}\fi

\bibitem[Aldahdooh et~al.(2021)Aldahdooh, Hamidouche, and Deforges]{aldahdooh2021reveal}
Ahmed Aldahdooh, Wassim Hamidouche, and Olivier Deforges.
\newblock Reveal of vision transformers robustness against adversarial attacks.
\newblock \emph{arXiv preprint arXiv:2106.03734}, 2021.

\bibitem[Bai et~al.(2021)Bai, Mei, Yuille, and Xie]{bai2021transformers}
Yutong Bai, Jieru Mei, Alan~L Yuille, and Cihang Xie.
\newblock Are transformers more robust than cnns?
\newblock \emph{Advances in neural information processing systems}, 34:\penalty0 26831--26843, 2021.

\bibitem[Bhojanapalli et~al.(2021)Bhojanapalli, Chakrabarti, Glasner, Li, Unterthiner, and Veit]{bhojanapalli2021understanding}
Srinadh Bhojanapalli, Ayan Chakrabarti, Daniel Glasner, Daliang Li, Thomas Unterthiner, and Andreas Veit.
\newblock Understanding robustness of transformers for image classification.
\newblock In \emph{Proceedings of the IEEE/CVF international conference on computer vision}, pages 10231--10241, 2021.

\bibitem[Brown(2020)]{brown2020language}
Tom~B Brown.
\newblock Language models are few-shot learners.
\newblock \emph{arXiv preprint arXiv:2005.14165}, 2020.

\bibitem[Carlini and Wagner(2017)]{carlini2017towards}
Nicholas Carlini and David Wagner.
\newblock Towards evaluating the robustness of neural networks.
\newblock In \emph{2017 ieee symposium on security and privacy (sp)}, pages 39--57. Ieee, 2017.

\bibitem[Carlini et~al.(2023)Carlini, Jagielski, Choquette-Choo, Paleka, Pearce, Anderson, Terzis, Thomas, and Tram{\`e}r]{carlini2023poisoning}
Nicholas Carlini, Matthew Jagielski, Christopher~A Choquette-Choo, Daniel Paleka, Will Pearce, Hyrum Anderson, Andreas Terzis, Kurt Thomas, and Florian Tram{\`e}r.
\newblock Poisoning web-scale training datasets is practical.
\newblock \emph{arXiv preprint arXiv:2302.10149}, 2023.

\bibitem[Caron et~al.(2021)Caron, Touvron, Misra, J{\'e}gou, Mairal, Bojanowski, and Joulin]{caron2021emerging}
Mathilde Caron, Hugo Touvron, Ishan Misra, Herv{\'e} J{\'e}gou, Julien Mairal, Piotr Bojanowski, and Armand Joulin.
\newblock Emerging properties in self-supervised vision transformers.
\newblock In \emph{Proceedings of the IEEE/CVF international conference on computer vision}, pages 9650--9660, 2021.

\bibitem[Devlin(2018)]{devlin2018bert}
Jacob Devlin.
\newblock Bert: Pre-training of deep bidirectional transformers for language understanding.
\newblock \emph{arXiv preprint arXiv:1810.04805}, 2018.

\bibitem[Dong et~al.(2018)Dong, Liao, Pang, Su, Zhu, Hu, and Li]{dong2018boosting}
Yinpeng Dong, Fangzhou Liao, Tianyu Pang, Hang Su, Jun Zhu, Xiaolin Hu, and Jianguo Li.
\newblock Boosting adversarial attacks with momentum.
\newblock In \emph{Proceedings of the IEEE conference on computer vision and pattern recognition}, pages 9185--9193, 2018.

\bibitem[Dosovitskiy et~al.(2020)Dosovitskiy, Beyer, Kolesnikov, Weissenborn, Zhai, Unterthiner, Dehghani, Minderer, Heigold, Gelly, et~al.]{dosovitskiy2020image}
Alexey Dosovitskiy, Lucas Beyer, Alexander Kolesnikov, Dirk Weissenborn, Xiaohua Zhai, Thomas Unterthiner, Mostafa Dehghani, Matthias Minderer, Georg Heigold, Sylvain Gelly, et~al.
\newblock An image is worth 16x16 words: Transformers for image recognition at scale.
\newblock \emph{arXiv preprint arXiv:2010.11929}, 2020.

\bibitem[Fort(2021{\natexlab{a}})]{Fort2021CLIPadversarial}
Stanislav Fort.
\newblock Adversarial examples for the openai clip in its zero-shot classification regime and their semantic generalization, 2021{\natexlab{a}}.

\bibitem[Fort(2021{\natexlab{b}})]{Fort2021CLIPadversarialstickers}
Stanislav Fort.
\newblock Pixels still beat text: Attacking the openai clip model with text patches and adversarial pixel perturbations, 2021{\natexlab{b}}.

\bibitem[Fu et~al.(2022)Fu, Zhang, Wu, Wan, and Lin]{fu2022patch}
Yonggan Fu, Shunyao Zhang, Shang Wu, Cheng Wan, and Yingyan Lin.
\newblock Patch-fool: Are vision transformers always robust against adversarial perturbations?
\newblock \emph{arXiv preprint arXiv:2203.08392}, 2022.

\bibitem[Goodfellow et~al.(2014)Goodfellow, Shlens, and Szegedy]{goodfellow2014explaining}
Ian~J Goodfellow, Jonathon Shlens, and Christian Szegedy.
\newblock Explaining and harnessing adversarial examples.
\newblock \emph{arXiv preprint arXiv:1412.6572}, 2014.

\bibitem[Gu et~al.(2022)Gu, Tresp, and Qin]{gu2022vision}
Jindong Gu, Volker Tresp, and Yao Qin.
\newblock Are vision transformers robust to patch perturbations?
\newblock In \emph{European Conference on Computer Vision}, pages 404--421. Springer, 2022.

\bibitem[He et~al.(2023)He, Jia, Liang, Lou, Liu, and Cao]{he2023sa}
Bangyan He, Xiaojun Jia, Siyuan Liang, Tianrui Lou, Yang Liu, and Xiaochun Cao.
\newblock Sa-attack: Improving adversarial transferability of vision-language pre-training models via self-augmentation.
\newblock \emph{arXiv preprint arXiv:2312.04913}, 2023.

\bibitem[Ilyas et~al.(2018)Ilyas, Engstrom, Athalye, and Lin]{ilyas2018black}
Andrew Ilyas, Logan Engstrom, Anish Athalye, and Jessy Lin.
\newblock Black-box adversarial attacks with limited queries and information.
\newblock In \emph{International conference on machine learning}, pages 2137--2146. PMLR, 2018.

\bibitem[Jia et~al.(2021)Jia, Yang, Xia, Chen, Parekh, Pham, Le, Duerig, and Song]{jia2021scaling}
Chao Jia, Yinfei Yang, Ye Xia, Yi-Ting Chen, Zarana Parekh, Hieu Pham, Quoc~V Le, Tom Duerig, and Claire Song.
\newblock Scaling up vision-language pretraining.
\newblock In \emph{International Conference on Machine Learning}, pages 4904--4916. PMLR, 2021.

\bibitem[Kim and Lee(2022)]{kim2022analyzing}
Gihyun Kim and Jong-Seok Lee.
\newblock Analyzing adversarial robustness of vision transformers against spatial and spectral attacks.
\newblock \emph{arXiv preprint arXiv:2208.09602}, 2022.

\bibitem[Kim et~al.(2024)Kim, Park, Kim, and Lee]{kim2024curved}
Juyeop Kim, Junha Park, Songkuk Kim, and Jong-Seok Lee.
\newblock Curved representation space of vision transformers.
\newblock In \emph{Proceedings of the AAAI Conference on Artificial Intelligence}, pages 13142--13150, 2024.

\bibitem[Kirillov et~al.(2023)Kirillov, Mintun, Ravi, Mao, Rolland, Gustafson, Xiao, Whitehead, Berg, Lo, et~al.]{kirillov2023segment}
Alexander Kirillov, Eric Mintun, Nikhila Ravi, Hanzi Mao, Chloe Rolland, Laura Gustafson, Tete Xiao, Spencer Whitehead, Alexander~C Berg, Wan-Yen Lo, et~al.
\newblock Segment anything.
\newblock In \emph{Proceedings of the IEEE/CVF International Conference on Computer Vision}, pages 4015--4026, 2023.

\bibitem[Kowalczuk et~al.(2024)Kowalczuk, Dubi{\'n}ski, Ghomi, Sui, Stein, Wu, Cresswell, Boenisch, and Dziedzic]{kowalczuk2024benchmarking}
Antoni Kowalczuk, Jan Dubi{\'n}ski, Atiyeh~Ashari Ghomi, Yi Sui, George Stein, Jiapeng Wu, Jesse~C Cresswell, Franziska Boenisch, and Adam Dziedzic.
\newblock Benchmarking robust self-supervised learning across diverse downstream tasks.
\newblock \emph{arXiv preprint arXiv:2407.12588}, 2024.

\bibitem[Kurakin et~al.(2018)Kurakin, Goodfellow, and Bengio]{kurakin2018adversarial}
Alexey Kurakin, Ian~J Goodfellow, and Samy Bengio.
\newblock Adversarial examples in the physical world.
\newblock In \emph{Artificial intelligence safety and security}, pages 99--112. Chapman and Hall/CRC, 2018.

\bibitem[Li et~al.(2021)Li, Selvaraju, Gotmare, Joty, Xiong, and Hoi]{li2021align}
Junnan Li, Ramprasaath Selvaraju, Akhilesh Gotmare, Shafiq Joty, Caiming Xiong, and Steven Chu~Hong Hoi.
\newblock Align before fuse: Vision and language representation learning with momentum distillation.
\newblock \emph{Advances in neural information processing systems}, 34:\penalty0 9694--9705, 2021.

\bibitem[Li et~al.(2023)Li, Li, Savarese, and Hoi]{li2023blip}
Junnan Li, Dongxu Li, Silvio Savarese, and Steven Hoi.
\newblock Blip-2: Bootstrapping language-image pre-training with frozen image encoders and large language models.
\newblock In \emph{International conference on machine learning}, pages 19730--19742. PMLR, 2023.

\bibitem[Lin et~al.(2015)Lin, Maire, Belongie, Bourdev, Girshick, Hays, Perona, Ramanan, Zitnick, and Dollár]{lin2015microsoft}
Tsung-Yi Lin, Michael Maire, Serge Belongie, Lubomir Bourdev, Ross Girshick, James Hays, Pietro Perona, Deva Ramanan, C.~Lawrence Zitnick, and Piotr Dollár.
\newblock Microsoft coco: Common objects in context, 2015.

\bibitem[Liu et~al.(2022)Liu, Tang, Liang, Gong, Wu, Liu, and Tao]{liu2022exploring}
Aishan Liu, Shiyu Tang, Siyuan Liang, Ruihao Gong, Boxi Wu, Xianglong Liu, and Dacheng Tao.
\newblock Exploring the relationship between architecture and adversarially robust generalization.
\newblock \emph{arXiv preprint arXiv:2209.14105}, 2022.

\bibitem[Loshchilov and Hutter(2017)]{loshchilov2017decoupled}
Ilya Loshchilov and Frank Hutter.
\newblock Decoupled weight decay regularization.
\newblock \emph{arXiv preprint arXiv:1711.05101}, 2017.

\bibitem[Lu et~al.(2023)Lu, Wang, Wang, Guan, Gao, and Zheng]{lu2023set}
Dong Lu, Zhiqiang Wang, Teng Wang, Weili Guan, Hongchang Gao, and Feng Zheng.
\newblock Set-level guidance attack: Boosting adversarial transferability of vision-language pre-training models.
\newblock In \emph{Proceedings of the IEEE/CVF International Conference on Computer Vision}, pages 102--111, 2023.

\bibitem[Madry et~al.(2017)Madry, Makelov, Schmidt, Tsipras, and Vladu]{pgd}
Aleksander Madry, Aleksandar Makelov, Ludwig Schmidt, Dimitris Tsipras, and Adrian Vladu.
\newblock Towards deep learning models resistant to adversarial attacks.
\newblock \emph{arXiv preprint arXiv:1706.06083}, 2017.

\bibitem[Mahmood et~al.(2021)Mahmood, Mahmood, and Van~Dijk]{mahmood2021robustness}
Kaleel Mahmood, Rigel Mahmood, and Marten Van~Dijk.
\newblock On the robustness of vision transformers to adversarial examples.
\newblock In \emph{Proceedings of the IEEE/CVF international conference on computer vision}, pages 7838--7847, 2021.

\bibitem[Moosavi-Dezfooli et~al.(2016)Moosavi-Dezfooli, Fawzi, and Frossard]{moosavi2016deepfool}
Seyed-Mohsen Moosavi-Dezfooli, Alhussein Fawzi, and Pascal Frossard.
\newblock Deepfool: a simple and accurate method to fool deep neural networks.
\newblock In \emph{Proceedings of the IEEE conference on computer vision and pattern recognition}, pages 2574--2582, 2016.

\bibitem[Naseer et~al.(2021)Naseer, Ranasinghe, Khan, Khan, and Porikli]{naseer2021improving}
Muzammal Naseer, Kanchana Ranasinghe, Salman Khan, Fahad~Shahbaz Khan, and Fatih Porikli.
\newblock On improving adversarial transferability of vision transformers.
\newblock \emph{arXiv preprint arXiv:2106.04169}, 2021.

\bibitem[Noever and Noever(2021)]{noever2021reading}
David~A Noever and Samantha E~Miller Noever.
\newblock Reading isn't believing: Adversarial attacks on multi-modal neurons.
\newblock \emph{arXiv preprint arXiv:2103.10480}, 2021.

\bibitem[Oquab et~al.(2023)Oquab, Darcet, Moutakanni, Vo, Szafraniec, Khalidov, Fernandez, Haziza, Massa, El-Nouby, et~al.]{oquab2023dinov2}
Maxime Oquab, Timoth{\'e}e Darcet, Th{\'e}o Moutakanni, Huy Vo, Marc Szafraniec, Vasil Khalidov, Pierre Fernandez, Daniel Haziza, Francisco Massa, Alaaeldin El-Nouby, et~al.
\newblock Dinov2: Learning robust visual features without supervision.
\newblock \emph{arXiv preprint arXiv:2304.07193}, 2023.

\bibitem[Radford et~al.(2021)Radford, Kim, Hallacy, Ramesh, Goh, Agarwal, Sastry, Askell, Mishkin, Clark, et~al.]{radford2021learning}
Alec Radford, Jong~Wook Kim, Chris Hallacy, Aditya Ramesh, Gabriel Goh, Sandhini Agarwal, Girish Sastry, Amanda Askell, Pamela Mishkin, Jack Clark, et~al.
\newblock Learning transferable visual models from natural language supervision.
\newblock In \emph{International conference on machine learning}, pages 8748--8763. PMLR, 2021.

\bibitem[Russakovsky et~al.(2015)Russakovsky, Deng, Su, Krause, Satheesh, Ma, Huang, Karpathy, Khosla, Bernstein, Berg, and Fei-Fei]{ILSVRC15}
Olga Russakovsky, Jia Deng, Hao Su, Jonathan Krause, Sanjeev Satheesh, Sean Ma, Zhiheng Huang, Andrej Karpathy, Aditya Khosla, Michael Bernstein, Alexander~C. Berg, and Li Fei-Fei.
\newblock {ImageNet Large Scale Visual Recognition Challenge}.
\newblock \emph{International Journal of Computer Vision (IJCV)}, 115\penalty0 (3):\penalty0 211--252, 2015.

\bibitem[Shao et~al.(2021)Shao, Shi, Yi, Chen, and Hsieh]{shao2021adversarial}
Rulin Shao, Zhouxing Shi, Jinfeng Yi, Pin-Yu Chen, and Cho-Jui Hsieh.
\newblock On the adversarial robustness of vision transformers.
\newblock \emph{arXiv preprint arXiv:2103.15670}, 2021.

\bibitem[Sharif~Razavian et~al.(2014)Sharif~Razavian, Azizpour, Sullivan, and Carlsson]{sharif2014cnn}
Ali Sharif~Razavian, Hossein Azizpour, Josephine Sullivan, and Stefan Carlsson.
\newblock Cnn features off-the-shelf: an astounding baseline for recognition.
\newblock In \emph{Proceedings of the IEEE conference on computer vision and pattern recognition workshops}, pages 806--813, 2014.

\bibitem[Silberman et~al.(2012)Silberman, Hoiem, Kohli, and Fergus]{silberman2012indoor}
Nathan Silberman, Derek Hoiem, Pushmeet Kohli, and Rob Fergus.
\newblock Indoor segmentation and support inference from rgbd images.
\newblock In \emph{Computer Vision--ECCV 2012: 12th European Conference on Computer Vision, Florence, Italy, October 7-13, 2012, Proceedings, Part V 12}, pages 746--760. Springer, 2012.

\bibitem[Uesato et~al.(2018)Uesato, O’donoghue, Kohli, and Oord]{uesato2018adversarial}
Jonathan Uesato, Brendan O’donoghue, Pushmeet Kohli, and Aaron Oord.
\newblock Adversarial risk and the dangers of evaluating against weak attacks.
\newblock In \emph{International conference on machine learning}, pages 5025--5034. PMLR, 2018.

\bibitem[Wang et~al.(2024)Wang, Jia, Liu, and Song]{wang2024benchmarking}
Chenguang Wang, Ruoxi Jia, Xin Liu, and Dawn Song.
\newblock Benchmarking zero-shot robustness of multimodal foundation models: A pilot study.
\newblock \emph{arXiv preprint arXiv:2403.10499}, 2024.

\bibitem[Wang et~al.(2023)Wang, Hu, Dong, and Hong]{wang2023exploring}
Youze Wang, Wenbo Hu, Yinpeng Dong, and Richang Hong.
\newblock Exploring transferability of multimodal adversarial samples for vision-language pre-training models with contrastive learning.
\newblock \emph{arXiv preprint arXiv:2308.12636}, 2023.

\bibitem[Wei et~al.(2022)Wei, Chen, Goldblum, Wu, Goldstein, and Jiang]{wei2022towards}
Zhipeng Wei, Jingjing Chen, Micah Goldblum, Zuxuan Wu, Tom Goldstein, and Yu-Gang Jiang.
\newblock Towards transferable adversarial attacks on vision transformers.
\newblock In \emph{Proceedings of the AAAI Conference on Artificial Intelligence}, pages 2668--2676, 2022.

\bibitem[Yang et~al.(2022)Yang, Duan, Tran, Xu, Chanda, Chen, Zeng, Chilimbi, and Huang]{yang2022vision}
Jinyu Yang, Jiali Duan, Son Tran, Yi Xu, Sampath Chanda, Liqun Chen, Belinda Zeng, Trishul Chilimbi, and Junzhou Huang.
\newblock Vision-language pre-training with triple contrastive learning.
\newblock In \emph{Proceedings of the IEEE/CVF Conference on Computer Vision and Pattern Recognition}, pages 15671--15680, 2022.

\bibitem[Young et~al.(2014)Young, Lai, Hodosh, and Hockenmaier]{young2014image}
Peter Young, Alice Lai, Micah Hodosh, and Julia Hockenmaier.
\newblock From image descriptions to visual denotations: New similarity metrics for semantic inference over event descriptions.
\newblock \emph{Transactions of the Association for Computational Linguistics}, 2:\penalty0 67--78, 2014.

\bibitem[Yuan et~al.(2021)Yuan, Hou, Jiang, Feng, Cheng, Wan, Xie, Kumar, Shi, Yu, et~al.]{yuan2021florence}
Lu Yuan, Qibin Hou, Zihang Jiang, Zhe Feng, Mingfei Cheng, Abner Wan, Jiadong Xie, Varun Kumar, Hongyu Shi, Dongdong Yu, et~al.
\newblock Florence: A new foundation model for computer vision.
\newblock \emph{arXiv preprint arXiv:2111.11432}, 2021.

\bibitem[Zhang et~al.(2022)Zhang, Yi, and Sang]{zhang2022towards}
Jiaming Zhang, Qi Yi, and Jitao Sang.
\newblock Towards adversarial attack on vision-language pre-training models.
\newblock In \emph{Proceedings of the 30th ACM International Conference on Multimedia}, pages 5005--5013, 2022.

\bibitem[Zhang et~al.(2023)Zhang, Huang, Wu, and Lyu]{zhang2023transferable}
Jianping Zhang, Yizhan Huang, Weibin Wu, and Michael~R Lyu.
\newblock Transferable adversarial attacks on vision transformers with token gradient regularization.
\newblock In \emph{Proceedings of the IEEE/CVF Conference on Computer Vision and Pattern Recognition}, pages 16415--16424, 2023.

\bibitem[Zhang et~al.(2024)Zhang, Huang, and Bai]{zhang2024universal}
Peng-Fei Zhang, Zi Huang, and Guangdong Bai.
\newblock Universal adversarial perturbations for vision-language pre-trained models.
\newblock In \emph{Proceedings of the 47th International ACM SIGIR Conference on Research and Development in Information Retrieval}, pages 862--871, 2024.

\bibitem[Zhou et~al.(2019)Zhou, Zhao, Puig, Xiao, Fidler, Barriuso, and Torralba]{zhou2019semantic}
Bolei Zhou, Hang Zhao, Xavier Puig, Tete Xiao, Sanja Fidler, Adela Barriuso, and Antonio Torralba.
\newblock Semantic understanding of scenes through the ade20k dataset.
\newblock \emph{International Journal of Computer Vision}, 127\penalty0 (3):\penalty0 302--321, 2019.

\end{thebibliography}
}

\end{document}